\documentclass[acmsmall,screen,nonacm]{acmart}

\usepackage{braket}
\usepackage{xcolor}
\usepackage{hyperref}
\usepackage{makecell}
\usepackage{graphicx}
\usepackage{amsmath,amsfonts,amsthm,bm,mathtools} 
\graphicspath{ {./figures/} }
\usepackage{caption}
\usepackage{subcaption}
\usepackage{multirow}

\newcommand{\m}[1]{\ensuremath{\mathrm{#1}}}

\begin{document}

\title{Quantum Annealing for Industry Applications: Introduction and Review }


\author{Sheir Yarkoni}
\affiliation{\institution{ Leiden University} \city{Leiden} \country{Netherlands}}

\author{Elena Raponi}
\email{e.raponi@liacs.leidenuniv.nl}
\orcid{0000-0001-6841-7409}
\affiliation{%
  \institution{Leiden University}
  \city{Leiden}
  \country{Netherlands}
}

\author{Thomas B\"ack}
\affiliation{\institution{Leiden University} \city{Leiden}\country{Netherlands}}

\author{Sebastian Schmitt}
\affiliation{\institution{Honda Research Institute Europe GmbH}\country{Germany}}
\email{Sebastian.Schmitt@honda-ri.de}
\orcid{0000-0001-7130-5483}

\renewcommand{\shortauthors}{Yarkoni et al.}

\begin{abstract}
Quantum annealing is a heuristic quantum optimization algorithm that can be used to solve combinatorial optimization problems. In recent years, advances in quantum technologies have enabled the development of small- and intermediate-scale quantum processors that implement the quantum annealing algorithm for programmable use. Specifically, quantum annealing processors produced by D-Wave Systems have been studied and tested extensively in both research and industrial settings across different disciplines. In this paper we provide a literature review of the theoretical motivations for quantum annealing as a heuristic quantum optimization algorithm, the software and hardware that is required to use such quantum processors, and the state-of-the-art applications and proofs-of-concepts that have been demonstrated using them. The goal of our review is to provide a centralized and condensed source regarding applications of quantum annealing technology. We identify the advantages, limitations, and potential of quantum annealing for both researchers and practitioners from various fields. 
\end{abstract}

\maketitle

\section{Introduction}
\label{sec:intro}

Quantum computing (QC) is a rapidly growing research field that promises a novel paradigm to solve challenging computational problems. It was first introduced in the early 1980s by physicist Paul Benioff~\cite{benioff_computer_1980,benioff_quantum_1982} and independently by Richard Feynman~\cite{feynman_simulating_1982}. These original proposals of quantum computers refer to the use of quantum phenomena such as superposition and entanglement to perform computation and simulate nature. In the decades since, significant advancement of both algorithms and hardware has been made that expanded the potential and scope of use of quantum computers. Seminal works from Deutsch~\cite{deutsch_quantum_1989} and Grover~\cite{grover} showed that quantum algorithms can provide an asymptotic speedup (meaning, by definition, non-constant speedup in the limit of large system sizes) relative to their classical counterparts. In some cases this can even result in exponentially faster algorithms, such as Shor's algorithm for factoring~\cite{shor} and the quantum Fourier transform algorithm~\cite{qft}. More recently, industrial efforts to build commercially-available QCs have emerged around the world, with vendors such as IBM~\cite{ibm65qubits}, Google~\cite{Arute2019}, Rigetti~\cite{rigetti}, D-Wave Systems~\cite{johnson_quantum_2011}, IonQ~\cite{ionq2,ionq} and others offering quantum computing services via cloud-based access. 

There are many different paradigms and hardware implementations that can be used to build quantum computers. The two leading paradigms are gate-based quantum computing and adiabatic quantum computation (AQC) \cite{farhi_AQC_2000,moritaMathQA2008,dasQAreview2008,albashAdiabaticQC2018,hauke_perspectives_2019}. In the gate model of QC the computation is performed by applying a sequence of unitary gates to a set of quantum bits (qubits), whose  states can be measured at the end of the computation~\cite{deutsch_quantum_1989}. In contrast, in AQC one prepares an initial many-qubit quantum state as the ground state of a simple Hamiltonian, to which an adiabatic time evolution is applied, changing the system to a final Hamiltonian whose ground state encodes the solution of the desired (optimization) problem. In principle, AQC has been shown to be polynomially equivalent to gate-based universal QC, as any quantum circuit can be represented as a time-dependent Hamiltonian with at most polynomial overhead~\cite{aharonov2008, biamonteACQ08}.

There are many possible technologies one could use to build functioning qubits, and with them quantum processing units (QPUs). The first such prototypes were based on nuclear magnetic resonance (NMR) technology as early as the 1990's~\cite{nmr1,nmr2}. Since then, other technologies have shown some promise in implementing QCs such as trapped ions~\cite{ionq,PRXQuantum.2.020343}, cold atoms~\cite{PhysRevLett.115.043003,kasper2021,Scholl2021}, photon-based quantum computers~\cite{xanadu}, Rydberg atoms~\cite{Scholl2021,ebadiRydbergMIS22}, and diamond cavities~\cite{nv}. There are many examples in literature of quantum annealing implemented in different architectures (particularly in atomic QPUs~\cite{Hauke2015,Gra2016,Glaetzle2017,Torggler2017,Qiu2020}). However, to date, the largest QCs use superconducting qubits. The technical motivations behind each technology (and their respective strengths and weaknesses) is beyond the scope of this review, but we refer the interested reader to the relevant citations above. 

An approach closely related to AQC is quantum annealing (QA)~\cite{moritaMathQA2008,mcgeoch_adiabatic_2014, hauke_perspectives_2019,dasQAreview2008,tanakaQuantumSpinGlasses2017}, proposed in 1988 by Apolloni et al. \cite{apolloni_numerical_1988}. QA can be viewed as a relaxation of the AQC model, where the conditions of adiabacity are not met, resulting in a heuristic variational quantum algorithm. Therefore, this approach can be used to find the ground state of Ising models~\cite{mccoy_two-dimensional_2013}, a known NP-hard task~\cite{barahona1982}. It is well-established in literature how to transform canonical NP-hard and NP-complete combinatorial optimization problems to forms suitable for quantum annealers~\cite{lucas_ising_2014, lucas_2018}.
These problems can be posed either in Ising form using a $\{-1, 1\}$ basis and spin variables, or as a quadratic unconstrained binary optimization (QUBO) problem using the $\{0, 1\}$ basis and binary variables. The two forms are equivalent and problems in one form can be easily represented in the other using a simple change of basis. As such, QA lends itself naturally to practitioners interested in using quantum algorithms to solve combinatorial optimization problems. While the question of whether QA can provide a quantum speed-up over classical approaches is still subject of a scientific debate (see, for example, \cite{hauke_perspectives_2019,mcgeoch_adiabatic_2014,troyerspeedup} and references therein), there is ongoing dedicated effort to use QA hardware to solve application-motivated problems. Numerous examples have been published of the use of QA to solve real-world problems in a variety of domains, particularly in the fields of optimization, scheduling, machine learning, and simulations of natural systems.
Here, a similar conclusion has to be drawn as, to the best of our knowledge, there is no industrial application where QA unquestionably outperforms classical heuristic algorithms. However, the technological advances of recent years and the potential that may still be realized in the near future fuel the hope that QA hardware devices may be competitive in some domains in the future. 

In QA, the solution to a problem is realized by going through several stages. Although the overall methodology shares many common points across different architectures, the details of each step in the process depend on the type of qubits used, the adiabatic protocol that is implemented, and other engineering considerations. Currently, D-Wave Systems annealers, which use superconducting qubits, are the most popularly used QA devices, so their functional logic is important for practitioners in this field to understand. Furthermore, there is a wide selection of papers from both research groups and industry evaluating the use of the D-Wave platform for various applications. Promising alternative approaches such as the Lechner-Hauke-Zoller (LHZ)~\cite{lhz} scheme or platforms utilizing cold atoms as developed by Pasqal~\cite{Scholl2021} or QuEra~\cite{ebadi2021} are currently also making massive strides in hardware development. However, while these atomic QPUs can implement adiabatic protocols based on Rydberg atom Hamiltonians (distinguished as \textit{analog mode} rather than quantum circuits which are \textit{digital mode}), they require a different paradigm to solve problems. Furthermore, the technology is currently at its early stages, and there are not many application-relevant works in this area. For an overview on potential applications of Rydberg atom quantum computers, see~\cite{QuEra_apps}. 

\begin{figure}
         \includegraphics[width=\textwidth]{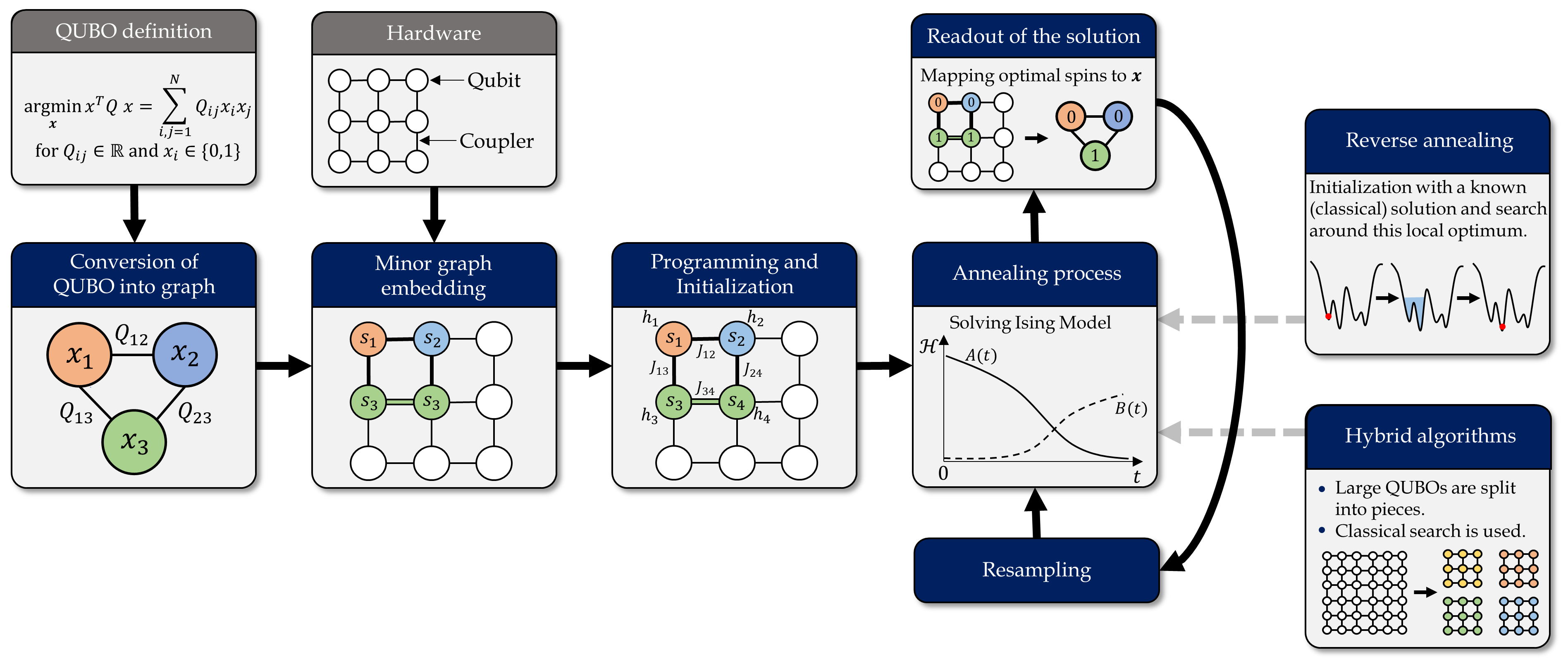}
         \caption{Visualization of a typical quantum algorithm workflow on a D-Wave quantum annealer. The image is an adapted and extended version of the workflow presented by \citet{goodrich_optimizing_2018}. }
         \label{fig:workflow}
\end{figure}

We now present the steps required to solve industry-relevant problems using QA as implemented by D-Wave Systems QPUs as a template. The concepts behind these steps will be explained in more detail in the upcoming sections, but the general process is as follows (which is also visualized in Fig.~\ref{fig:workflow}):
\begin{itemize}
    \item \textbf{Definition of a QUBO formulation and graph representation.} QUBO has become the standard input format for quantum annealers, to which the real-world application problems are converted. After definition, the QUBO is directly converted into a logical graph, where each node represents a variable and each edge denotes the interaction term between a pair of variables. 
    \item \textbf{Minor-embedding.} When translating the logical interaction graph onto the physical hardware graph of the quantum processing unit (QPU), it is necessary to select sets of physical qubits to represent a single logical node and to identify the couplings between the physical qubits to realize the correct interactions  between the logical variables. 
    \item \textbf{Programming.} Programming the quantum annealer requires setting the parameters that define the problem to be solved, also called final Hamiltonian. This involves setting the weights for each qubit bias (to control the magnetic field acting on the qubit) and coupler strength (to control the interaction between qubits). 
    \item \textbf{Initialization.} After programming, the spin configuration of the QPU is initialized as the lowest energy configuration of an easy-to-implement initial Hamiltonian, where the qubits are placed in an equal superposition of all possible states.
    \item \textbf{Annealing process.} In this step, the Ising model is solved. The system transitions from the initial to the final Hamiltonian according to predefined annealing functions in an attempt to minimize energy. This step can also be the core of a hybrid approach where the inner loop of the calculation is performed by a quantum processor. Annealing can also be combined with a reverse annealing phase, which initializes the quantum annealer with a known (classical) solution and searches the state space around this local optimum.
    \item \textbf{Readout of the solution.} At the end of the annealing phase, the qubits are in an eigenstate, or superposition of eigenstates, in the computational basis, where each eigenstate represents a possible minimum of the final Hamiltonian. The individual spin values of the qubits are read out and stored externally representing a candidate solution to the original problem.
    \item \textbf{Resampling.} Quantum annealing is a heuristic, which means that there is only ever a non-zero probability that the computation results in a ground state of the system. Therefore, the anneal-readout cycle is repeated many times per input to acquire multiple candidate solutions.
\end{itemize}

In this work we give an overview of the current state-of-the-art industrial applications and proofs-of-concepts that have been demonstrated using QA hardware. 
We focus on industrial problems that have been addressed using QA, and also provide insights on the limitations imposed by utilizing these methods. We present results that mainly use D-Wave quantum annealers, which reflects the fact that these are currently the only commercially available large-scale quantum annealers. The purpose of this review is to give a guide to practitioners whether their problems might be addressable using QA technology. We discuss potential benefits and limitations regarding the type of problem, the manageable problem sizes, the type of constraints to be included and potentially other relevant aspects for practical applications. We highlight the most prominent examples in existing literature, and attempt to bridge the gap between the fundamental quantum physics governing QA and application development. The ultimate goal of this work is to concisely present the most pertinent information required by newcomers to the field of quantum annealing to properly understand both the potential and limitations of QA for industry-related problems. 

The rest of this paper is structured as follows. 
In Section \ref{sec:basics} we introduce the foundations of quantum annealing. Starting from the central theoretical ideas, we go through the definition of QUBOs and explain how to embed constraints in QUBOs formulations. We illustrate how binary encoding schemes can be used to represent discrete and continuous variables, respectively. We introduce the hardware implementation of quantum processors, and describe how to embed the QUBO/Ising models into a physically realized QPU by minor-embedding.
We address the problem of noise and discuss the mitigation strategies proposed so far. 
We describe currently available software and show how classical and quantum resources can leverage their complementary strengths by introducing the concept of hybrid quantum-classical algorithms.
The role of time in quantum annealing is discussed in detail. Section \ref{sec:real-world} reviews a selected number of state-of-the-art applications and empirical studies involving QA. We focus on five relevant areas: traffic flow optimization, scheduling and logistics, quantum chemistry, physics and biology, quantum machine learning, and finance.
We then elaborate on these ideas and provide some concluding remarks on the potential and limitations of QA for real-world applications.
In Section \ref{sec:lessons}, we summarize the key lessons learned at the end of our journey through quantum annealing.
Finally, in Section \ref{sec:conclusions}, we draw our conclusions.

\section{Foundations of Quantum Annealing}

\label{sec:basics}
The purpose of this section is to briefly explain the physics behind adiabatic quantum computation (AQC) and quantum annealing (QA), with sufficient detail to give a newcomer to the field a basic understanding of the foundations of the approach. For a more in-depth analysis of the physics behind QA, we refer the reader to Refs.~\cite{hauke_perspectives_2019,tanakaQuantumSpinGlasses2017, albashAdiabaticQC2018}.

\subsection{Theoretical background}
\label{sec:theory}

The motivation for QA is found in the \textit{adiabatic theorem} \cite{born_beweis_1928,moritaMathQA2008}, which states the following: a quantum system in its ground state will remain in the ground state, provided the Hamiltonian governing the dynamics changes sufficiently slowly. This condition of ``sufficiently slowly'' is dictated by the smallest energy difference between the ground state and the first excited state exhibited during the system's evolution (also known as the minimum gap). 
This can be utilized for computation by preparing a system in the ground state of an easy to solve initial Hamiltonian and then switching the Hamiltonian sufficiently slow to a complex Hamiltonian. 
By the adiabatic theorem, the system will remain in the ground state and consequently  one ends up with the ground state, i.e., the solution of the complicated Hamiltonian.   
Applying this to Ising Hamiltonians, where it is NP-hard to find the minimal energy configuration, one can theoretically use such a quantum approach to solve problems in NP (but not necessarily in polynomial time). Furthermore, the adiabatic theorem can be used to simulate any quantum circuit, making it a universal quantum computing paradigm~\cite{aharonov2008}. This model of quantum computing is known as adiabatic quantum computing (AQC) or adiabatic quantum optimization (AQO).

The quantum algorithm is started by preparing the system in the ground state of a Hamiltonian, which is known and easy to prepare, denoted as $\mathcal{H}_i$ (also known as the \textit{initial Hamiltonian}). Then, the system is slowly changed such that the contribution of $\mathcal{H}_{i}$ is slowly reduced while the magnitude of a final (also known as \textit{target}) Hamiltonian, denoted $\mathcal{H}_f$ is increased (using time parameter $t$): 
\begin{equation}
	\mathcal{H}(t) = A(t)\mathcal{H}_i + B(t)\mathcal{H}_f,
	\label{eq:ising_model}
\end{equation}
Time $t \in [0,T_a]$ and the annealing schedule is defined by the monotonic functions $A(t)$ and $B(t)$ with $A(t=0)=1$, $B(t=0)=0$ and $A(t=T_a)=0$, $B(T_a)=1$. 
The gradual transition from $\mathcal{H}_i$ to $\mathcal{H}_f$ induces an evolution from the initial (ground) state $|\Psi(t=0)\rangle$ to the ground state $|\Psi(t=T_a)\rangle$ of the final Hamiltonian $\mathcal{H}_f$, where $|\Psi(t)\rangle$ represents the time-dependent wavefunction in the Dirac bra-ket notation (for details, see, for example, \cite{hauke_perspectives_2019}).
Typically, a transverse field in $x$-direction is used as the initial Hamiltonian, 
\begin{equation}
\mathcal{H}_{i} =  \sum_{i\in V} \sigma^x_i,
\end{equation}
where $\sigma^x_i$ is the $x$-Pauli matrix acting on the $i$-th qubit. $\sigma^x$ induces flips of the $\sigma^z$-basis states as $\sigma^x|\pm\rangle=  |\mp\rangle$ and its eigenstates are $|x\pm\rangle=\frac1{\sqrt{2}}\big(|+\rangle\pm|-\rangle\big)$, which are equal amplitude superpositions of the $\sigma^z$ eigenstates. Thus, initially, the system is prepared in the ground state to $\mathcal{H}_{i}$, which is given by $ |\Phi(t=0)\rangle = | x-,\dots, x- \rangle$. The final Hamiltonian is represented by:
\begin{equation}
	\mathcal{H}_{f} = \sum_{i\in V} h_i \sigma^z_i + \sum_{(i,j)\in E} J_{ij} \sigma^z_i \sigma^z_j.
	\label{eq:ising_ham}
\end{equation}

Here, $V$ is the set of vertices of the graph $G(V,E)$ representing the lattice sites where qubits  are located, $E$ the set of edges of the graph connecting the qubits,  $J_{ij}=J_{ji}$ are the symmetric interaction strength of the qubits $i$ and $j$ connected by an edge  and $h_i$ is the on-site energy (local field) of qubit $i$. 
For any given time $t$, the  Hamiltonian $\mathcal{H}(t)$ of Eq.~\eqref{eq:ising_model} is the transverse-field Ising Hamiltonian, which is well studied  in literature \citep{stinchcombe_ising_1973}. As the magnitude of $\mathcal{H}_i$ decreases, the quantum dynamics of the qubits slows down until at $t=T_a$ we obtain a purely classical system, where the final states of the qubits are measured. The classical Ising model is obtained by replacing the quantum $\sigma^z$ Pauli operators by classical spin variables (all other terms are the same as in Eq.~\eqref{eq:ising_ham}):

\begin{equation}
    \mathcal{H}_{\m{Ising}} = \sum_{i} h_i s_i + \sum_{ij} J_{ij} s_i s_j.
	\label{eq:ising_ham_classic}
\end{equation}

Despite the theoretical guarantees of the adiabatic theorem, the condition of adiabaticity is difficult to maintain in practice. Specifically, the fact that open quantum systems are susceptible to background noise and thermal fluctuation means that the quantum system could be kicked out of its ground state by the environment. 
Furthermore, the condition of ``sufficiently slowly'' is hard to fulfil, as the allowed rate of change is determined by the low energy states and the gaps in the spectrum of $\mathcal{H}(t)$, which is not known \textit{a priori}.
QA can be seen as a relaxation of AQC, where the annealing schedule, i.e., the time  to switch from $\mathcal{H}_{i}$ to $\mathcal{H}_{f}$, is determined heuristically and thus the strict adiabatic conditions are not guaranteed.  
Therefore, instead of a deterministic quantum algorithm, we instead obtain a \textit{heuristic optimization algorithm}, where a non-zero probability of remaining in the ground state is maintained during annealing, i.e., during the complete time evolution. 

The resulting QA heuristic is rather similar in spirit to the well-known classical simulated annealing approach to optimization~\cite{kirkpatrick_optimization_1983}. Simulated annealing exploits thermal fluctuations for the exploration of the search space. During the optimization procedure, the strength of the thermal fluctuations is gradually reduced and for sufficiently slow annealing schedule global optima can be found. QA can be viewed to implement a very similar logic but where the thermal fluctuations are  replaced by quantum fluctuations. The term proportional to $\mathcal{H}_i$ in Eq.\ \eqref{eq:ising_model} induces disorder with respect to the eigenstates of the Ising Hamiltonian in Eq.\ \eqref{eq:ising_ham} and introduces energetic variations to the qubits during the annealing process. During the annealing, the strength of the quantum fluctuation, i.e., the weight of $\mathcal{H}_{i}$ is gradually reduced. The hope that quantum mechanical effects lead to an advantage of QA over classical algorithms is related to the conjectured role of quantum superposition, tunneling, and entanglement~\cite{albashAdiabaticQC2018}.
The basic reasoning is that quantum superposition and tunneling allow for a direct transition between states even if there is a high energy barrier between them~\cite{bookQuantumTunneling}, which leads to the ability of a search algorithm to escape local minima by tunneling through energetic barriers. 
In simple experiments, it has been demonstrated rigorously that superposition and entanglement are indeed present throughout the annealing process~\cite{LantingEntanglement2019}. However, it is not clear how these properties translate to an important role in the optimization process in large ensembles of qubits (see, for example, Refs.,\ \cite{hauke_perspectives_2019, muthukrishnan_tunneling_2016,katzgraberQuantumSpeedup2015, denchevBenefitTunneling2016} and references therein for more detailed discussions).

Currently, state-of-the-art QA hardware can only implement a subset of the protocols necessary that allow for universal quantum computation. Nonetheless, it has been shown that allowing increased control of the annealing path can significantly improve the quality of results obtained while still in the absence of universality. Specifically, the introduction of \emph{non-stoquastic}\footnote{A Hamiltonian is called stoquastic in a given basis if it only has real and non-positive off-diagonal matrix elements. Therefore, a non-stoquastic Hamiltonian has positive or complex-valued off-diagonal elements. Stoquastic Hamiltonians can be simulated classically as no sign problem occurs.}
Hamiltonian terms result in increased degrees of freedom during the computation, giving hope for increasing the power of QA implementations~\cite{nonstoquastic_ozfidan,vinci_non-stoquastic_2017}. Further evidence is given by the fact that it is a known hard problem to simulate such Hamiltonians classically, and is therefore a bottleneck in quantum Monte Carlo procedures~\cite{Marvian2019,guptaSignProblemStoquastic2020}. It has been shown that implementing these terms to the driver Hamiltonian $\mathcal{H}_i$ can result in mitigation of prohibitive phase transitions at the avoided crossing~\cite{nonstoquastic_nishimori}. This mitigation can result in exponential speedup of the quantum annealing algorithm with non-stoquastic $\mathcal{H}_i$~\cite{nonstoquastic_choi}. 
However, the situation is currently not fully clear, as there are also indications that adding non-stoquastic terms to the Hamiltonian reduces the energy gap between the ground and first excited state, which makes QA less robust~\cite{CrossonDesigningNonStoquastic2020}.

Another relatively novel protocol implemented in recent QPUs is the \emph{reverse annealing} procedure~\cite{noauthor_reverse_2017}. Contrary to regular (forward) annealing, the Hamiltonian starts as a purely classical one, with the transverse field turned off (i.e., $\mathcal{H}_f$ being the only component). Then, each qubit's classical state can be set independently by the user as a starting point. The transverse field is then slowly raised, inducing a partial mixing of classical and quantum components to the Hamiltonian, and then lowered again (the relative magnitude of $\mathcal{H}_i$ is a free parameter). It has been shown that by using this novel protocol, reverse annealing can be useful in ways that strictly forward annealing cannot. For instance, one could seed a classical candidate solution to $\mathcal{H}_f$ and use reverse annealing to perform semi-local neighborhood search~\cite{noauthor_reverse_2017}. Or alternatively, using the reverse annealing as the Markovian step in Markov chain Monte Carlo~\cite{King2018}, or as a mutation operator in genetic algorithms~\cite{QAGA}. It has also been shown in multiple applications that reverse annealing may yield solutions that strictly forward annealing cannot, due to a fundamentally different annealing path being used~\cite{nbmf_ra2,ikeda_application_2019,Venturelli2019,grant_benchmarking_2021}.

From a practitioner's standpoint, QA can overall be viewed as a black-box heuristic optimization algorithm that attempts to find the minimum of a classical Ising spin  system. In the Ising formulation, optimization problems are transformed to the spin  model in Eq.~\eqref{eq:ising_ham_classic}. It has been shown that finding the ground state configuration of this spin system is in the worst case NP-hard~\cite{barahona1982}, and is therefore of practical interest to the field of optimization. Therefore, the role of the practitioner is to find a suitable Ising representation of the optimization problem of interest, which is then used as the final Hamiltonian in the QA algorithm, Eq.~\eqref{eq:ising_ham}. 

\subsection{Quadratic unconstrained binary optimization formulation}
\label{sec:qubo}
The Ising model can be easily transformed into various other forms, such as the unconstrained binary quadratic program (UBQP)~\cite{glover_diversification-driven_2010}, the undirected graphical model~\cite{madigan_bayesian_1993}, and the quadratic unconstrained binary optimization (QUBO) problem~\cite{lucas_ising_2014}. Most commonly, either the QUBO or Ising model is used to construct problems that can be solved by QA. The QUBO form is equivalent to the Ising model through a simple change of basis ($s \mapsto 2x-1$). 
In this representation, the goal is to find the configurations of $N$ binary variables $x_i \in \{0,1\}$, $i=1,...,N$ that minimize the objective function
\begin{equation}
\label{eq:qubo}
	\mathrm{Obj}(\mathbf{x}, Q) = \mathbf{x}^T \cdot Q \cdot \mathbf{x} = \sum_{i,j=1}^N  Q_{ij}\, x_i \, x_j\,,
\end{equation}
where $\mathbf{x}$ is the vector of the $N$ binary variables and $Q$ is the symmetric matrix defining interaction terms between the variables. 

\subsection{Higher-order binary optimization problems}
In general, it is possible to address higher-order polynomials in binary optimization with a variety of techniques (see, e.g., Ref.\ \cite{hauke_perspectives_2019}).
These are called either polynomial unconstrained binary optimization (PUBO) or higher-order unconstrained binary optimization (HUBO) problems. It is relatively straightforward to reformulate  problems with local $k$-th order interactions, also called $k$-local terms, with $k>2$ to QUBO, but this incurs a polynomial overhead in the number of variables leading to an undesirable scaling of the problem size~\cite{valiante_scaling_overhead_2020,ishikawaReduction11}
Consider a simple three-body (i.e.\ 3-local) term, $x_1x_2x_3$. Obviously, this is not a QUBO. However, we can introduce a new variable, $y_1 \coloneqq x_1x_2$, which we can substitute into our expression to obtain $y_1x_3$, which is a QUBO. However, this requires us to add a penalty term to the QUBO to constrain the new variable. Specifically, $y_1$ needs to follow the Boolean expression: $y_1 \Leftrightarrow x_1 \land x_2$. This expression has a known QUBO formulation with a minimum when the expression is satisfied, $x_1x_2 - 2(x_1 + x_2)y_1 +3y_1$. This acts as the constraint on the new $y_1$, and we obtain the final representation:
\begin{equation}
    x_1x_2x_3 = y_1x_3 + \lambda \big(x_1x_2 - 2(x_1 + x_2)y_1 + 3y_1 \big),
\end{equation}
where $\lambda>1$ is a penalty scalar ensuring the validity of the constraint. It is worth noting that this procedure can be used recursively to replace any high-order terms with new variables, thus reducing arbitrary polynomials to QUBO (more techniques and examples can be found here~\cite{dwaveDocs_hubo}). In practice, since such techniques introduce a polynomial number of new variables to the QUBO, such problem representations quickly become intractable. 

One alternative paradigm to implement higher-order QUBO terms is the ParityQC architecture for LHZ encoding. In general, it has been shown that one could implement $k$-local terms with a $k$-dimensional lattice with the parity architecture~\cite{lhz}. With this technique one could implement all-to-all connectivity of a $k$-local Hamiltonian consisting of $N$ logical qubits using $N^k$ physical qubits. More recently, a further generalization has shown that a specific transformation exists (now known as the \textit{parity transformation}) which maps non-zero terms of an arbitrary $k$-local Hamiltonian to a single qubit in a 2D parity architecture. Thus, the total number of qubits required to encode an arbitrary Hamiltonian (and therefore arbitrary binary optimization problems) is linear in the number of non-zero terms in the Hamiltonian. However, for a Hamiltonian of $N$ spins and $(k \leq N)$-local terms, the number of non-zero terms is at most $\sum_{i=1}^{k} {N \choose i}$.

\subsection{Constrained optimization problems}
\label{sec:constrained_opt}

So far, we have discussed the use of quantum annealing purely in the context of unconstrained binary optimization. However, real-world applications typically involve other, more general forms of optimization problems, which may also be addressed using quantum annealing, albeit not natively. We review some of the prominent classes of non-binary and constrained optimization and their respective implementation, along with some relevant literature. 

It is important to note that constraints on the binary variables cannot be directly enforced in the QUBO/Ising formulation. Once the Hamiltonian is programmed, it is immutable and the annealing procedure determines some minimal energy configuration, where conditions or constraints on the binary variables cannot be applied. As a consequence, all constraints must therefore be enforced by contributions to the cost function in the Hamiltonian. A common approach to solving this problem is by implementing penalty terms with large penalty factors. 

The inclusion of linear equality constraints with integer coefficients is straight forward. Consider, for example, a simple one-hot constraint over 
three binary variables $(x_1, x_2, x_3)$, where exactly one variable must be set to one while the others need to be zero, i.e.,
\begin{equation}
     x_1 + x_2 + x_3 = 1.
\end{equation}

To ensure that such a constraint is obeyed by our QUBO formulation, we can use the following equivalent quadratic representation of the constraint,
\begin{equation}
    \lambda \left(x_1 + x_2 + x_3 - 1\right)^2 = 0,
    \label{eq:one_hot}
\end{equation}
and add a corresponding quadratic term $ \lambda \left(x_1 + x_2 + x_3 - 1\right)^2$ to the cost function. The minimum of this contribution is zero only when exactly one of the variables is 1 and the rest are 0, and it is larger than zero for any other configuration.  Therefore,  when adding such a term to the QUBO cost function, the contribution is vanishing only when the one-hot constraint is not violated. 
In this fashion, the inclusion of linear equality constraints $P(\mathbf{x})=\sum_ia_i x_i-b=0$ is realizable in a straight forward fashion by adding a penalty term
\begin{equation}
   \lambda \left(\sum_ia_i x_i-b\right)^2\equiv\lambda \,P(\mathbf{x})^2
    \label{eq:equal_constriant}
\end{equation}
to the cost function.

Similarly, problems with linear inequality constraints can also be expressed as penalty terms, with the addition of auxiliary slack variables. Consider the following inequality for three binary variables $(x_1, x_2, x_3)$:

\begin{equation}
    x_1 + x_2 + x_3 \leq 2.
\end{equation}

As before, we must transform this inequality to its suitable QUBO form. We start by adding auxiliary variables such that the total slack is accounted for: in this case, two additional binary variables $(y_1, y_2)$ are sufficient. We notice that the equality constraint 
\begin{equation}
x_1 + x_2 + x_3 = 2 - y_1 - y_2,    
\end{equation}
is equivalent to the original inequality constraint, since it can be fulfilled  for all values of the binary variables $x_i$ satisfying the original inequality by choosing some appropriate values for the slack variables. For values violating the original inequality constraint, the equation can never be fulfilled regardless of the values of the slack variables. 
The equality constraint can be written in the equivalent quadratic form and added as a penalty term to the QUBO cost function,
 \begin{equation}
    \lambda \left( x_1 + x_2 + x_3 + y_1 + y_2 - 2\right) ^2.
\end{equation}

More general linear inequality constraints $P(\mathbf{x})=\sum_ia_ix_i-b\leq 0$ with positive integer coefficients $a_i, b\in\mathbb{N}^+_0$ can be handled in this fashion by adding a penalty term
\begin{equation}
  \lambda \left(\sum_ia_i x_i+\sum_j^W w_j y_j-b\right)^2\equiv\lambda \,P(\mathbf{x})^2
    \label{eq:inequal_constriant}
\end{equation}
to the cost function. The number of slack variables $W$ and their coefficient $w_j$ can be derived from the coefficients $a_i$ and $b$, where at most $W=b$ slack variables are needed. The more general case with arbitrary integer coefficients, including negative coefficients, can also be handled but the number of slack variables increases to $W=b+\sum_{a_i<0}|a_i|$ in the worst case. The most general case with arbitrary real-valued coefficients cannot be handled in a straight forward manner with binary slack variables.

In this way, canonical NP-complete and NP-hard problems with linear inequalities can be transformed to QUBOs and Ising models~\cite{lucas_ising_2014}. However, this transformation comes at an increased cost in the number of optimization variables, the exact number of which depends on the exact problem being solved. For integer and binary problems this increase is usually polynomial in the number of original problem variables, which in turn increases the number of fully-connected components that need to be embedded. 
Notice that quadratic and higher order polynomial constraints cannot not be included in this fashion efficiently, as the quadratic penalty functions leads  to even higher order interactions in the cost function, which incur additional overhead in auxiliary binary variables.  More general non-linear constraints cannot be included in the QUBO formulation using penalty functions. 

In summary, the full QUBO cost function including constraints takes the form
\begin{align}
\label{eq:qubo_penalty}
    C(\mathbf{x})&=\mathrm{Obj}(\mathbf{x},Q)+\sum_{\text{constraints } k} \lambda_k P_k(\mathbf{x})^2
\end{align}
where $\mathrm{Obj}(\mathbf{x}, Q)$ is the original cost function of the  problem formulation of Eq.~\eqref{eq:qubo} and the $P_k(\mathbf{x})$ are linear functions in the binary variables representing the constraints, which are satisfied when $P_k(\mathbf{x})=0$.
However, we still need to ensure that it is never energetically favorable to minimize the objective function with a violated constraint.
This is the role of the $\lambda_k$ in Eq.~\eqref{eq:qubo_penalty}: the penalty, i.e., the cost of violating a single constraint should be dominating and therefore, should be greater than any other reduction in objective function value. 
In classical computing, this can be resolved by setting $\lambda_k$ to an arbitrarily large value. 
However, with quantum annealing hardware this value has to be chosen with care, due to the limited dynamic range of the possible coupling strengths and pre-factors. The larger this value, the less resolution is effectively realizable in the low energy-level regime of $\mathcal{H}_f$, respectively $\mathrm{Obj}(\mathbf{x},Q)$. Large penalty factors result in extreme amounts of distortion in the ground state distribution due to noise. Therefore, it is important to consider setting $\lambda_k$ such that the lowest possible value that maintains correctness of the constraints is used. While there is no universal way to set these parameters, simple rules-of-thumb may be found either algebraically or empirically~\cite{neukart_traffic_2017,srp}. In the case of purely constraint-based optimization problems such as constraint satisfaction problem (CSP) or boolean satisfiability (SAT) problems, no $\lambda$ is needed when summing independent constraints, since the minimum of each constraint, as well as the global optimum is at a cost function value of zero, $C(\mathbf{x}_{\m{solution}})=0$. Furthermore, it becomes trivial to verify whether a candidate solution is a satisfying statement for CSP/SAT by checking the objective value given the solution (finding such a solution is still NP-hard). Previous work has shown interesting academic results using SAT problems to construct Bloom-like filters using quantum annealers~\cite{douglass_constructing_2015}. Furthermore, it has been shown how to design optimal penalty terms for constraint satisfaction problems to minimize the error for quantum annealers~\cite{bian_circuit_fault}.

Another way of handling constraints for the QA approach was proposed more recently \cite{ohzeki_breaking_2020,yuHubbardStratQA2021}, which is very similar to Lagrangian relaxation in traditional combinatorial optimization (see, for example, Ref.~\cite{korteCombinatorialOptBook2021}). The approach starts by formulating the constraints as quadratic penalty terms to the QUBO cost function as described above. 
By considering the partition function of the QUBO cost function and employing a Hubbard-Stratonovich transformation for the quadratic constraints $P_k^2$ followed by a Wick rotation in the auxiliary variables, an alternative cost function can be derived (see the original works \cite{ohzeki_breaking_2020,yuHubbardStratQA2021} for details). 
This alternative cost function is linear in the constraints at the expense of additional auxiliary real-valued variables $\nu_k\in\mathbb{R}$, one for each constraint, and takes the form:
\begin{align}
    C'(\mathbf{x},\boldsymbol{\nu})&=\mathrm{Obj}(\mathbf{x},Q)-\sum_k\frac{\nu_k^2}{2\lambda_k}-\sum_{k} \nu_k P_k(\mathbf{x})\:.
\end{align}
Finding a solution that minimizes the original cost function $\mathrm{Obj}(\mathbf{x})$ and fulfills the constraints $P_k(\mathbf{x})=0$ is equivalent to finding a saddle point of the alternative cost function $C'(\mathbf{x},\boldsymbol{\nu})$ by calculating $\min_{\mathbf{x}}\max_{\boldsymbol{\nu}}C'(\mathbf{x},\boldsymbol{\nu})$.
Ohzeki \cite{ohzeki_breaking_2020} proposed an iterative scheme, where a gradient ascent is performed with respect to the variables $\boldsymbol{\nu}$, and in each step  the minimal energy configuration for the binary variables $\mathbf{x}$ is found using QA.
So instead of solving one QUBO on the QA device, a whole series of QUBOs has to be solved, one for each proposed set of auxiliary variables, which incurs a substantial computational overhead. On the other hand, the advantage is that the constraints are not included as quadratic but linear terms in the QUBO, which reduces the overhead in binary variables related to the minor embedding of quadratic coupling terms. As a consequence larger problem instances  involving more binary variables can be approached. An additional benefit of the proposed approach is that constraints that are quadratic in the binary variables can directly be incorporated into the QUBO formulation.
It should be noted that this method of treating constraints has only been proposed recently and few application problems have been considered. Apart from some technicalities, which are not so clear from the original works, the effectiveness and robustness of the proposed method has to be evaluated on more problems in the future in order to arrive at a solid understanding of the benefits and  shortcomings of the method. 

It is important to note that, in theory, it is also possible to add constraints in the solution of a problem by implementing specific driver Hamiltonians to the evolution of a quantum system. Thus, the system is constrained to a subspace of the total Hilbert space during its evolution as specified by the driver. However, to be maximally effective, error-correction may be required, as it has been shown that in open quantum systems (meaning, in the presence of noise) such constraints are not guaranteed to be obeyed~\cite{streif_driver}. The LHZ architecture is one such paradigm which can benefit from constraints being implemented in the driver Hamiltonian of a quantum annealing protocol. While for all-to-all connected Hamiltonians the enumeration of these constraints is straightforward, in practice graphs can be much sparser, and therefore require just-in-time problem-dependent compilation in order to be mapped correctly. A detailed explanation of how such considerations arise and are handled can be found in~\cite{ParityQC_constraints}.

\subsection{Discrete variables}
\label{sec:discrete}

Discrete variables are of vital importance in many optimization problems, and have been investigated in the context of quantum annealing, for example in scheduling problems \cite{venturelli_quantum_2016, rieffel_case_2015} and graph colouring \cite{titiloye_quantum_2011,TitiloyeGraphColoring_HybridQA2011, marxGraphColoring2004}. Each categorical variable can be encoded by using multiple qubits to represent a logical integer variable subject to a single constraint~\cite{tamuraEncoding2021,karimiIntegerEncoding2019,harris_digital_2012}. Binary encoding, analogous to classical binary encoding, is efficient in the number of qubits in the sense that one can encode $d$ discrete states using only $\lceil\m{log}_2(d)\rceil$ qubits. However, the binary encoding is not used much in practice for application problems, since the interactions necessary to enforce the validity of the encoding as well as the realization of the couplings between logical variables  are rather complicated to implement with the binary encoding.  
Additionally, this encoding has been investigated in previous works and has been shown to be detrimental to the QPU's ability to find ground states~\cite{Chancellor2019, domain_wall_performance}.

The one-hot encoding case is a standard technique where a logical variable with $d$ possible states is represented by $d$ qubits. Each of those qubits corresponds to one possible value of the discrete variable if set to the up-state $\ket{+}$, whereas all other qubits need to be in the down-state $\ket{-}$. 
The corresponding constraint is typically implemented as quadratic interaction terms between all $d$ qubits. For limited-topology hardware this represents a drawback due to the embedding overhead, which causes noise to take a dominant role in a QPU's ability to find ground states~\cite{noise_SK}.

A recently developed alternative is the domain-wall encoding, which can encode discrete variables with one fewer qubit per variable, i.e., $d-1$ qubits for $d$ states. Additionally, it does not require interactions between all qubits that represent one variable, but instead only a linear coupling. It has been shown to be more efficient in different test problems~\cite{Chancellor2019, domain_wall_performance,berwaldDomainWallEncoding2021}. 
In particular, a recent work showed that domain wall encoding leads to preferable dynamics during the annealing process, which in turn leads to an increased probability of finding feasible solutions, among other benefits~\cite{berwaldDomainWallEncoding2021}.

\subsection{Continuous variables}
\label{sec:continuous}
Representing continuous numbers using binary variables is not well-suited for quantum annealing due to precision issues with analog QPUs. Much like in classical computing, these numbers are represented by binary encoding schemes. A single floating-point variable $\Tilde{x}$ with $N$ bits of precision would be encoded as:

\begin{equation}
    \Tilde{x} = c \sum_{i=0}^{N-1} b^i x_i\,,
\end{equation}
where $x_i\in\{0,1\}$ are the binary variables, $b>1$ defines the logarithmic resolution and $c\in\mathbb{R}$ is a scale factor defining the range of the floating point variable.  
Such a continuous encoding is not well suited for QA approaches as the unavoidable noise of each binary variable on the QPU has profoundly different influence on the variable depending on its prefactor, which are exponentially spaced.  
Additionally, the number of qubits needed to encode such continuous variables needs to be rather large.
For the very few practical applications that utilize such an encoding, the precision requirements are the most detrimental factor in QPU performance~\cite{Ottaviani2018,srp}. This means that to solve high-precision problems necessarily means that (in the worst case) the precision level is below the noise threshold for QPUs.

\subsection{Hardware architecture for quantum annealers}
\label{sec:hardware}

There exist multiple initiatives to build quantum annealing hardware in industry. In Japan, NEC corporation recently announced\footnote{\url{https://www.nec.com/en/press/202203/global_20220317_01.html}} their use of parametron qubits~\cite{NEC_parametron_original} to build a unit cell for a quantum annealer using LHZ encoding~\cite{NEC_LHZ}\footnote{As part of a New Energy and Industrial Technology Development Organization (NEDO) project, \url{https://www.nedo.go.jp/english/activities/activities_ZZJP_100123.html}}. In Europe, Qilimanjaro aims to build high-coherence processors to enable both quantum annealing and gate-model quantum computers~\cite{qilimanjaro}. Currently, D-Wave Systems is the largest-scale and best-known of these initiatives, launching their first QA hardware product line in 2011~\cite{johnson_quantum_2011} and offering regularly updated hardware platforms ever since.

The core of an annealing QPU is an array of qubits, each of which realizes a simple two-level (e.g., spin-$\frac{1}{2}$) quantum system. D-Wave Systems utilize superconducting flux qubits to build their QPUs~\cite{harris_experimental_2010}. For a full technical description of the quantum processors and the physics behind each component, we refer the reader to~\cite{johnson_quantum_2011}. In principle, annealers could be built from almost any of the qubit technologies mentioned in Section~\ref{sec:intro}, although some exhibit certain properties that are amenable to quantum annealing (e.g., transmon qubits~\cite{leibTransmonQA2016}).

One property that all superconducting quantum annealing QPUs have in common (especially in the absence of error-correction) is that the physical layout of the qubits is fixed, typically referred to as a \emph{hardware graph} and denoted by $U$. This naming convention comes from the graph-theoretical description of the QPU, where each qubit (coupler) is represented by a node (edge) in an undirected graph. The exact topology of the hardware graph dictates the structure of graphs and possible $\mathcal{H}_f$ that can be represented natively by the qubits' connectivity. The QPU topology is an engineering artifact arising from the design choices for the QPU. In general, due to engineering difficulties, connectivity between qubits comes at the cost of the total number of qubits in the QPU~\footnote{An exception to this is atomic QPUs. In cold atom processors, the atoms (and therefore the qubits) can be arranged in various layouts as a starting point for computation. With trapped ions, an all-to-all connectivity is possible.}.  

Due to various technological limitations in manufacturing, calibration, or other anomalies, some qubits and couplers in D-Wave QPUs may be defective and hence not programmable. The percentage of qubits and couplers that remain functional once fully exposed to users is referred to as the \emph{hardware yield}. For D-Wave QPUs, the qubit yield is typically around $97\%$ for current processors~\cite{dwaveQPUprops}. The graph representing the functional qubits and couplers is referred to as the QPU's \emph{working graph}, and is therefore a subgraph of a full hardware graph. Depending where (and how many) the missing qubits and couplers occur, these effects limit the size and structures of problems that can be solved on the QPU~\cite{Lobe2021}.

\begin{figure}[h!]
     \centering
     \begin{subfigure}[b]{0.46\textwidth}
         \centering
         \includegraphics[width=\textwidth]{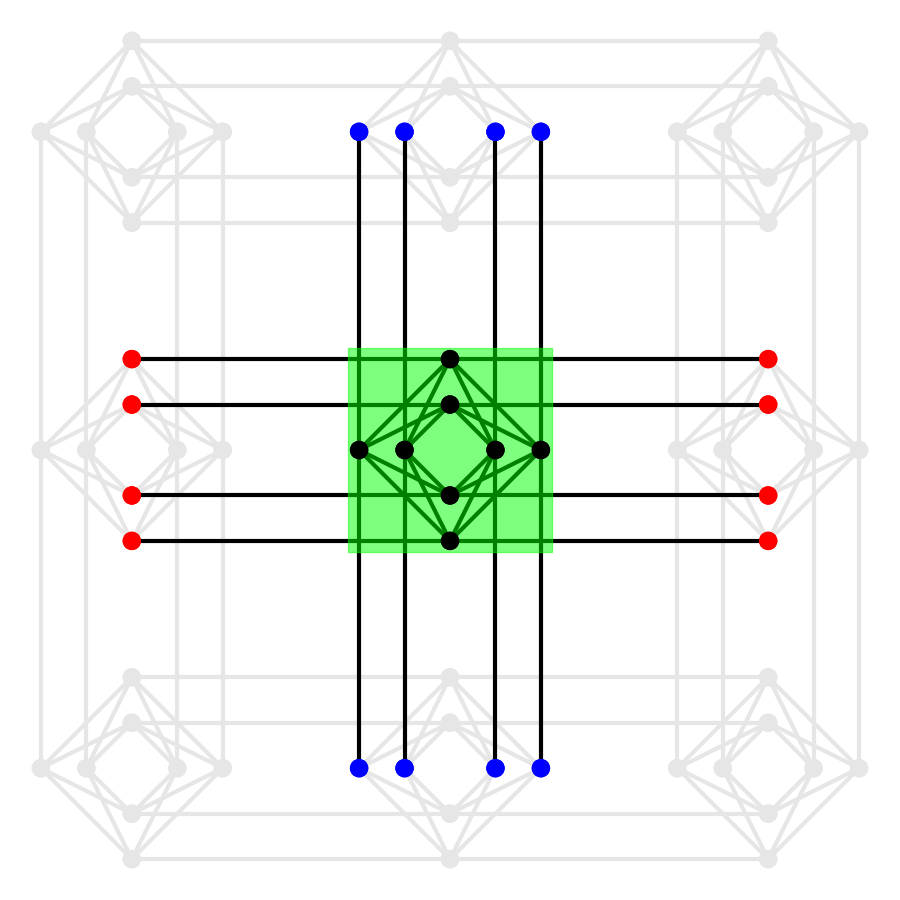}
         \caption{         }
         \label{fig:chimera}
     \end{subfigure}
     \hfill
     \begin{subfigure}[b]{0.48\textwidth}
         \centering
         \includegraphics[width=\textwidth]{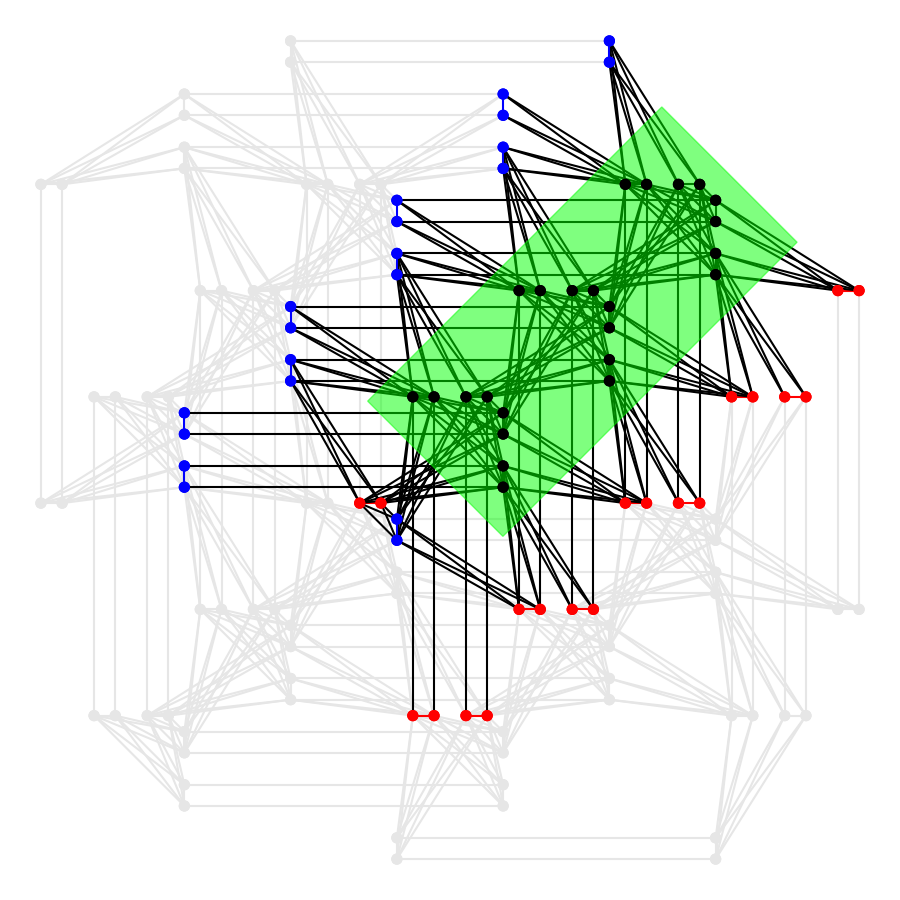}
         \caption{ }
         \label{fig:pegasus}
     \end{subfigure}
        \caption{Graphical representations of the topology graphs of two D-Wave hardware architectures. Each qubit is represented by a node in the graph, and every coupler is an edge between nodes.
        (a) Chimera graph: the highlighted green square is a $K_{4,4}$ unit cell. Blue and red nodes are vertically and horizontally connected qubits to the highlighted unit cell. 
        (b) Pegasus graph: the highlighted region represents the equivalent unit cell, composed of $K_4$ and $K_{6,6}$ subgraphs. Blue and red nodes here represent vertical and horizontal pairs of qubits connected to the highlighted cell. Each qubit has at most degree 15 (explained in detail in  \url{https://docs.dwavesys.com/docs/latest/c_gs_4.html}). Both figures were made using the D-Wave Ocean Python package~\cite{dwave_ocean}.    
        }
        \label{fig:topologies}
\end{figure}

The specific topologies of the D-Wave 2000Q and earlier generations of QPUs are referred to as \emph{Chimera} topology. These graphs are composed of a 2D lattice of small complete bipartite graphs (each called a tile, or unit cell). Typically, these graphs are denoted as $C_X$ (where $X$ is the length of one side of the square lattice, for a total of $X^2$ tiles), and each tile is in itself a $K_{N,N}$ bipartite graph (full connectivity between left and right partitions). The D-Wave 2000Q has a size of $C_{16}$ and $K_{4,4}$ tiles. Each qubit in the Chimera topology has six couplers, where four couplers are inside the unit cell and two are to different unit cells. A diagram showcasing this topology is in Fig.~\ref{fig:chimera}. Due to their relatively sparse connectivity, Chimera topology QPUs have been found to not be conducive to quantum speedups and advantages for problems formulated directly on Chimera graphs (also known as \textit{native problems})~\cite{glassy}. The latest generation of D-Wave QPUs have a novel topology, called \emph{Pegasus}, which differs in two major aspect from the Chimera graph: first, the graph degree is 15 (meaning each qubit is connected to 15 neighbors), and secondly, the hardware graph is not bipartite. Instances of the Pegasus topology that contain $N\times N$ unit cells are referred to as $P_N$ and consist of $24N(N-1)$ qubits. This results in a significantly more complex structure, allowing for denser graph structures of $\mathcal{H}_f$ to be solved by the QPUs, and hence more difficult optimization problems (a technical description comparing Chimera and Pegasus can be found in Ref.~\cite{pegasus}). For a diagram of a Pegasus graph, see Fig.~\ref{fig:pegasus}.

\subsection{Minor-embedding techniques for hardware graphs}
\label{sec:minor-embedding}

It is evident that not all interesting optimization problems defined by the interaction graph $G$ specified in $\mathcal{H}_f$ can be natively expressed in the hardware topologies $U$, i.e., the Chimera or Pegasus graphs. However, it is possible to reformulate any arbitrarily-structured $\mathcal{H}_f$ into a new hardware-compatible $\mathcal{H}'_f$ via the technique of graph \emph{minor-embedding}, a well-studied problem in graph theory (see, e.g.\ Ref.~\cite{diestelGraphTheory}). This process produces a mapping between one graph to another such that the relevant topological properties of the original graph are preserved. For QA, this involves encoding variables as \emph{chains} of multiple physical qubits on a QPU such that they act as a single logical qubit~\cite{Choi2008,choi_minor-embedding_2011}. In Ref.~\cite{pegasus}, the authors explore the various chain requirements when embedding certain graph families in both Chimera and Pegasus graphs. 

There exist several difficulties when using minor-embedding techniques. First, deciding whether a graph can be minor-embedded into another is a known NP-complete problem, and so polynomial-time heuristics are used in practice~\cite{embedding,Boothby2016}. Secondly, the use of embeddings requires an additional set of constraints to be imposed on the qubits representing a logical qubit (the magnitude of which is called the \emph{chain strength}). These enforce that the minimum energy of $\mathcal{H}'_f$ is obtained only when all physical qubits representing a logical qubit are in the same state. It has been shown that the magnitude of the minimum chain strength increases with the degree of the graph of $\mathcal{H}_f$~\cite{choi_minor-embedding_2011, noise_SK,paintshop}, which can create distortions in the resulting Hamiltonian $\mathcal{H}'_f$ due to noise. Generally, determining the best suited chain strength coupling is nontrivial task and still subject of current research, see, e.g.~\cite{fangChainStrength2020}.

\begin{figure}[h]
         \includegraphics[width=.7\textwidth]{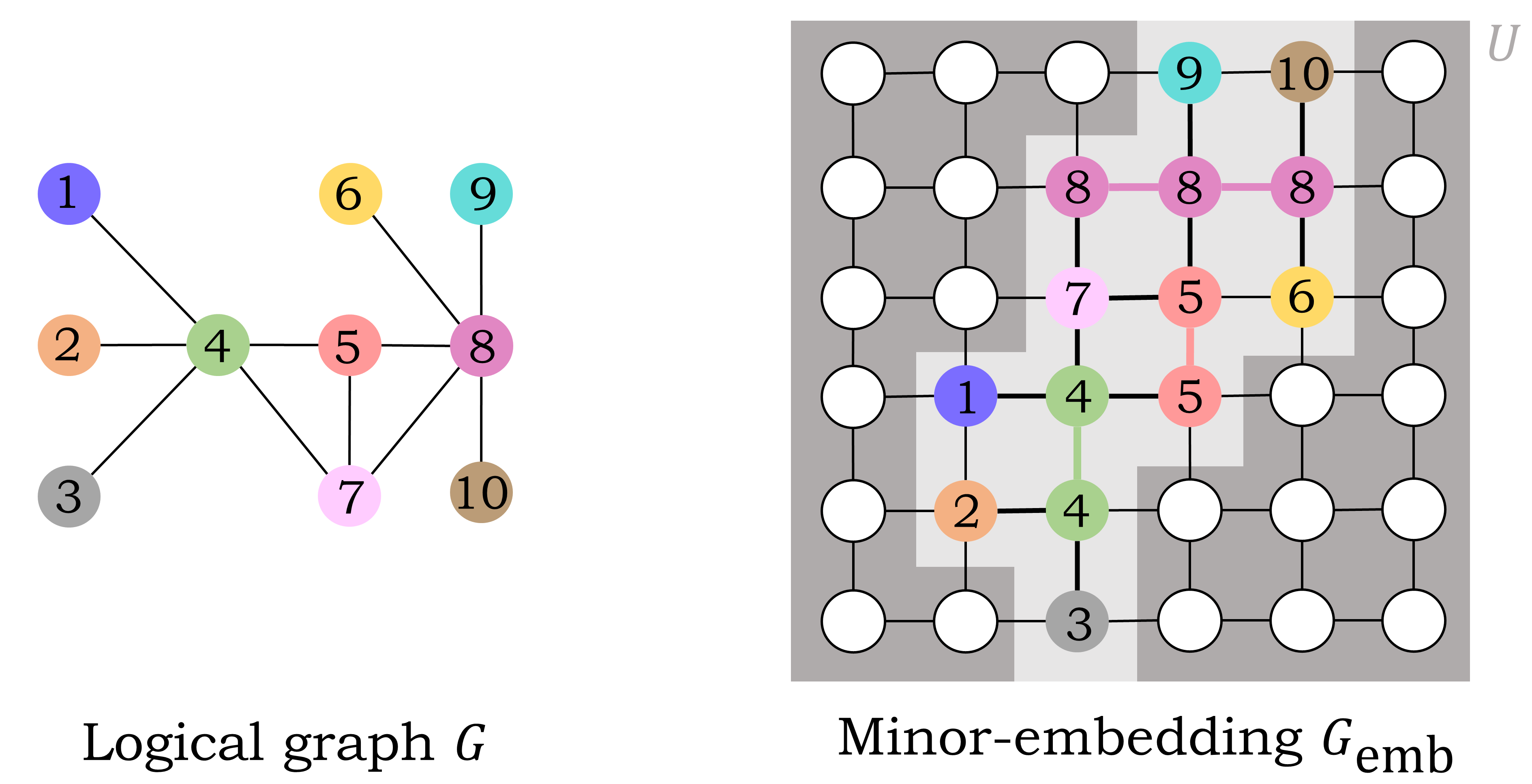}
         \caption{Logical graph $G$ and its minor-embedding $G_{\text{emb}}$ in a general square lattice $U$.
         Each node of $G$ is mapped to one or more nodes in $U$. Each group of nodes representing one logical node, i.e., each chain, is highlighted with a different label and color. Adapted from~\cite{Choi2008}.}
         \label{fig:embedding}
\end{figure}

An illustrative example of the minor embedding procedure and the necessary grouping of physical to logical qubits is depicted in Figure \ref{fig:embedding}. 
It can be observed, that the logical nodes $4$ and $8$ have degree $5$, which is larger degrees than the maximal degree $4$ of the hardware graph, which is assumed to be a simple square lattice for illustrative purposes. In the example, two physical nodes are used to represent the logical node $4$, i.e., form a chain. Also, the triangular coupling of the nodes $4$, $5$ and $7$ is incommensurate with the square lattice, and therefore the additional chain representing node $5$ has to be introduced.
Three physical nodes are required to represent the logical node $8$, since it has more than four connections as well as being part of a triangular coupling.

Various methods have been developed to find efficient minor-embeddings in various scenarios. \citet{embedding} proposed a heuristic algorithm for finding a minor embedding $G_{\m{emb}}$ of graph $G$ to the hardware topology graph $U$ that is practical for sparse $U$ and $G$ with hundreds of vertices. Previously, \citet{robertson_graph_1995} had proposed exact algorithms with the same aim, but these are impractical for graphs with more than tens of vertices. 
The success of a heuristic to find minor-embeddings heavily depends on the fact that, if $G$ is significantly smaller than $U$, then there are probably a large number of distinct minor-embeddings $G_\m{emb}$ in $U$, and their number grows exponentially with the number of vertices of $U$. 
Since the working graph $U$ may be different in each QPU, \citet{embedding} presented a heuristic algorithm where both $G$ and $U$ are given as an input. 
The heuristic algorithm aims to either minimize the maximal chain length (the maximal number of physical qubits representing any logical qubit), or the total embedding overhead (the total number of used physical qubits in $U$).

\citet{klymko_adiabatic_2012} were the first to provide a polynomial-time algorithm for generating large clique minors in subgraphs of a hardware graph. 
Via tree decomposition techniques, they develop an algorithm taking $O(N^4)$ time for a $C_{N,4}$ Chimera topology. This algorithm was one of the first embedding techniques constructed specifically to be resilient in the face of missing qubits in the hardware graph $U$.  Later, \citet{Boothby2016} defined a combinatorial class of \textit{native clique minors} in Chimera graphs with vertex images of uniform and near minimal size, and provide a polynomial-time algorithm that finds a maximum native clique minor in a given induced subgraph of a Chimera graph in only $O(N^3)$ time.

\citet{Zaribafiyan2017} proposed a systematic, deterministic, and scalable embedding algorithm for embedding the Cartesian product of two complete graphs into D-Wave Systems’ Chimera hardware graph. 
They defined an application-specific algorithm that exploits both the intrinsic structure of combinatorial optimization problems and the structure of the quantum annealer, defined by the Chimera topology.
Their Cartesian product of complete graphs (CPCG) embedding algorithm showed ability in obtaining high-quality embeddings on ideal Chimera chips with equal-length chains by using far fewer physical qubits, advantageous running time scaling and cost-effectiveness in scaling up for larger chip architectures over more-general heuristic methods. However, they did not explore optimal modifications to the method to handle inoperable qubits and couplers. In this context, Lobe et al.~\cite{Lobe2021} investigated such effects by embedding complete graphs on incomplete Chimera graphs. The authors introduced a method of formulating the embedding problem as an integer linear problem (ILP), which is fixed-parameter tractable in the number of missing qubits in the working graph, and were able to embed larger complete graphs than heuristic methods.

Although well established algorithms exist, the minor-embedding process is by itself a non-trivial problem, since it is the dominant costly step time-wise and the largest detractor on performance for QA.  
Depending on the problem and the hardware connectivity, some minor embedding algorithms might be inefficient or even fail to find a valid embedding.  
Typically, the number of physical qubits needed to  formulate a specific problem with only pairwise interactions on the Chimera topology of the actual hardware scales quadratically with the number of logical qubits of the problem, $N_\m{physical} \sim N^2_\m{logical}$.
For problems where true three-body and higher-order interaction terms have to be included in the Hamiltonian, as for instance it is naturally the case in the quantum chemistry domain, the scaling is much worse, typically exponential in the number of variables \cite{streif_solving_2019,valiante_scaling_overhead_2020}.

Recently, the increase in the time-to-solution due to the embedding overhead was compared between several embedding schemes~\cite{koenz_AnnealingEmbeddingOverhead_2021}. The  authors investigated the embedding of problems with fully connected graphs, i.e., with all-to-all couplings, 
onto the 2D planar square lattice graph, the Chimera graph of the actual D-Wave hardware and the recently proposed Lechner-Hauke-Zoller (LHZ) scheme~\cite{lhz}. All embedding schemes induced unfavorable exponential overhead in the time-to-solution with increasing problem size, underlining the necessity to investigate novel approaches to QA. When approaching practical application problems, D-Wave provides various tools and guidance for minor embedding in their documentation, see~\cite{dwaveDocs}.

\subsection{Noise and mitigation strategies}
\label{sec:noise}

The disordered, interacting, time-dependent and open nature of the many-body Ising Hamiltonian realizable in actual QA hardware systems makes it difficult to draw universal conclusions about the power of programmable QA~\cite{zagoskin_how_2014}.
However, when considering QA as a relaxation of AQC implemented in hardware, some general statements can be made in the face of noise. As typical of a metaheuristic, decoherence errors in QA have the effect of reducing the probability of obtaining global minima. However, it has been shown how despite the manifestation of finite-temperature effects in the samples obtained from a QA processor, this does not preclude the samples from being useful in practice~\cite{temperature}. Even in the case of decoherence, the annealing process is still completed and a candidate solution to $\mathcal{H}_f$ is obtained. It has been shown in practice that annealers still perform heuristic optimization in the absence of quantum coherence and entanglement~\cite{denchevBenefitTunneling2016,incoherent_tunnelling}. Furthermore, QA is somewhat resilient against dephasing errors, as the relative phases between qubits do not change the eigenvalue in the computational basis, and therefore the problem being solved. Dephasing effects ultimately affect the probability of obtaining global optima by changing the annealing path during the annealing process. Without full programmability of the annealing path it is hard to quantify the full effect of such errors.

Dickson et al.~\cite{Dickson2013} performed the first experimental demonstration of the effect of thermal noise on D-Wave QA processors~\cite{harris_experimental_2010, johnson_quantum_2011}. In this context, robustness against environmental noise was defined as the ability of an open QA system to yield the correct solution with acceptable probability within a time comparable to the closed-system adiabatic timescale. The authors examined this by constructing a 16-qubit toy problem designed to have a small energy gap between the lowest two energy levels. It was demonstrated that in the limit of weak coupling to the environment (i.e., relatively low levels of thermal noise), annealing non-adiabatically across the minimum gap did not hinder QA, but rather enhanced its performance. This was demonstrated by manually raising the operating temperature of the chamber in which the QPU was operating, thus allowing for a controllable amount of noise to perturb the system. The specific point demonstrated was that thermal noise does not necessarily inhibit obtaining ground states at the end of the annealing, even with small minimum gaps and non-adiabatic conditions. However, the caveat remains that the annealing times need to be sufficiently long, and the system must be only weakly coupled to the environment\footnote{Due to thermal relaxation, in the worst case the time to reach equilibrium in an open quantum system can grow  orders of magnitude with respect to the time needed in a closed system.}.

Despite the relative robustness of QA in terms of ground state probabilities, noise can also directly perturb problem formulations in the analog system, i.e., $\mathcal{H}_f$. The Hamiltonian parameters are subject to noise arising from a variety of sources, such as low-frequency noise, digital-to-analog discretization errors, and more. For the purposes of this work, we present a summary of the noise model in D-Wave quantum annealers, the sources of which are collectively known as Integrated Control Errors (ICE)~\cite{dwave_noise_description}. In D-Wave QA, the perturbation of the problem Hamiltonian due to noise can be modeled by a shift in the Hamiltonian terms as follows:
\begin{equation}
    \mathcal{H}^\delta_f = \sum_{i=1}^{N}\left(h_i + \delta h_i\right)s_i + \sum_{i=1}^{N}\sum_{j=1}^{N} \left(J_{ij}+\delta J_{ij}\right)s_is_j.
    \label{eq:noise}
\end{equation}
The specific values of the shifts are both hardware and problem dependent, but their magnitudes are roughly of the order of $|\delta h_i| \approx 0.05$ and $|\delta J_{ij}| \approx 0.02$ (in Ising units). It is important to note that the indices of~\eqref{eq:noise} do not necessarily strictly obey the QPU topology. Noise induced by next-nearest-neighbor effects results in what is known as \emph{ghost couplings}, where noise in physically unconnected qubits can cause effective couplings (and therefore correlations) through intermediary qubits and couplers. For a full description of the noise sources (and their measurements) in D-Wave QPUs we refer the reader to~\cite{dwave_noise_description}. It should be noted that a higher connectivity QPU with a denser hardware graph may cause noise aggregation on the original problem Hamiltonian. As shown by Willsch et al.~\cite{willsch_benchmarking_2022}, the increased connectivity of D-Wave's Pegasus topology can lead to a reduction in the success probability as compared to the Chimera topology when addressing sparsely connected problems. In this case, the results suggest that the unused couplers introduce additional noise. 

In the context of perturbed problem Hamiltonians, Venturelli et al.~\cite{noise_SK} compared QA and simulated thermal annealing to measure the effects of disorder in ground state distributions on the Sherrington-Kirkpatrick Model (SKM) with couplings randomized from a bimodal distribution of values $\pm1$ \cite{sherrington_solvable_1975}.
The two key findings from this work in regards to noise are as follows. Firstly, it was evident that the ferromagnetic couplings needed to encode chains representing  logical qubits on the annealer needed to increase as $\sqrt{N}$ (up to logarithmic corrections) in order to properly represent the SKM for larger sizes.
Secondly, that static noise (i.e., thermal noise whose timescale is much slower than the annealing dynamics) causes a reshuffling of the energy levels in the problem Hamiltonian. It was ultimately observed that the performance of simulated thermal annealing was similarly degraded in the face of such ground state disorder. However, when considering same-magnitude noise/disorder, it was observed that quantum annealing had a limited scaling advantage for this dataset. This further gives evidence that well-separated eigenstates are particularly important for the success of the quantum annealing algorithm. 

There are multiple ways to mitigate some of the effects of noise in quantum annealing. The most common technique is \emph{spin-reversal transforms}, which are used to effectively average over the asymmetric final Hamiltonian distortion due to noise. This is done by creating random spin vectors $S = \{-1, 1\}^N$ and recalculating a new $\bar{\mathcal{H}}_f$ subject to this shift, where $\bar{h}_i = h_i \cdot S(i)$, and $\bar{J}_{ij} = J_{ij} \cdot S(i)S(j)$. It is simple to show that these transformations do not change the distribution of ground states in the system, only the signs of a subset of terms. This new Hamiltonian $\bar{\mathcal{H}}_f$ is then submitted to the annealer. To recover the samples subject to the original Hamiltonian $\mathcal{H}_f$, we simply multiply the solutions by the vector $S(i)$. Typically, many such transforms are used when solving a single $\mathcal{H}_f$.  While this method is effective in reducing the error due to static noise, each new $\bar{\mathcal{H}}_f$ results in another programming cycle, increasing the total wall-clock time of using QA. Furthermore, it has been shown that increasing the number of transforms has diminishing returns for a fixed number of samples~\cite{temperature}. 

With its third generation quantum processors, D-Wave Systems introduced the possibility to control one of the annealing terms for each qubit individually. Specifically, it is possible to delay or advance the annealing start time of each qubit independently, within some predefined ranges. This feature is called \emph{anneal offsets}. The motivation behind this feature is that each qubit may experience a slow-down in dynamics at a different point in the annealing process, defined by the magnitude of $\mathcal{H}_f$ (known as the ``freeze-out'' effect). A naive example is embedded problems, which require chains of qubits with ferromagnetic couplings. Longer chains have a lower effective tunneling energy and may freeze in a fixed state earlier in the annealing process, called the freeze-out point. It has been shown that when this point is early in the annealing process, the resulting sample distributions are distorted by noise effects~\cite{Marshall_thermalization}. Anneal offsets can be used to shift the freeze-out time of each chain (or qubit) to mitigate this, and improve the probability of finding the ground-state.
Andriyash et al.~\cite{factoring_anneal_offsets} highlight the potential of the anneal offsets feature and addressed the problem of integer factoring, where given an input number $p$ one tries to find the integers $a$ and $b$ such that $p=a\, b$. 
They addressed the problem by executing a multiplication circuit backwards, i.e., the circuit has inputs $a$ and $b$ and achieves an optimal state that minimizes the optimization problem when $a \, b = p$. 
Employing the anneal offsets, they showed that the computation can be about 1000 times faster than without the offsets, reducing the solution time from $10^1s$ to $10^{-2}s$.
The same technique was also applied to 2-SAT problems with 12 Boolean variables \cite{hsu_quantum_2018}.
Annealing offsets have been shown to mitigate first-order phase transitions in simple problems that could be solved exactly, exponentially increasing performance compared to SA \citep{susa_exponential_2018}. 
Yarkoni et al.~\cite{Yarkoni2019} investigated the use of anneal offsets for random instances of the maximum independent set problem. They used the covariance matrix adaptation evolutionary strategy (CMA-ES) \cite{hansen_cma_2006-2} to heuristically optimize the annealing offset of the qubit chains, i.e., the $\Delta t$ of advance or delay for each qubit in physical time. Therefore, the functions $A(t)$ and $B(t)$ in Eq.~\eqref{eq:ising_model} are in principle different for each qubit, i.e., $A(t_i)$ and $B(t_i)$. They showed that by tuning the annealing offsets, quantum annealing can improve by an order of magnitude in ground state probability, and is able to also find better solutions than without annealing offsets. They observed that the improvement increases with problem size until a certain point, after which the gains mostly disappear due to the increase in variables to be optimized and the consequently harder optimization problem to be solves. Similarly, \citet{Adame_AnnealOffsets_2020} investigated classes of problems in particularly difficult regimes for D-Wave QPUs, with some instances showing success probabilities as low as $10^{-7}$. The authors employed inhomogeneous annealing offsets that depend on the local connectivity of each qubit and show that time-to-solution can be reduced by factors on the order of $10^3$ to $10^7$.

Alternative schemes to suppress errors in QA have been proposed that utilize ancillary qubits in the problem Hamiltonian~\cite{Pudenz2014}. The role of these extra qubits is to enforce penalties that act as analogues to classical bit-flip error mitigation strategies. Logical variables are encoded redundantly with multiple physical qubits, and classical states are recovered by a simple majority vote. While this procedure incurs a polynomial overhead in the number of qubits required to describe a problem, the authors show numerically that adding such penalty terms can increase the minimum gap, giving hope that QA may experience a speedup using these methods. 

Overall, it is clear that current state-of-the-art QA does not have as clear or established error-correction methodologies as gate-model QC does. Furthermore, it is clear that error-correction in QA requires fundamentally novel paradigms to be implemented. Nonetheless, ongoing research in both hardware development (additional control features, or alternative drivers) as well as software (spin reversals, tuning of annealing parameters) implies that QA may still significantly benefit from error-correction (and suppression).

\subsection{Software and hybrid algorithms}
\label{sec:hybrid}

Despite the impressive advancement of quantum hardware in recent years, development of high-level software to address quantum annealers has been slow. Many different aspects of using QA still require classical software, such as hardware control, formulation of Ising models/QUBOs, APIs, etc. Specifically for QA hardware access, the majority of the relevant software tools are owned, operated and developed only by D-Wave Systems.

D-Wave Systems maintains a set of publicly available repositories for application development in Python, named Ocean~\cite{dwave_ocean}. We briefly describe the most relevant ones that are relevant for practitioners. Constructing simple Ising/QUBO problems is done using the classes available in $\texttt{dimod}$. To submit these problems to D-Wave QPUs and solvers, D-Wave provides abstracted composite layers for both embedded and topology-specific problems in $\texttt{dwave-system}$. These composites also expose to the user the necessary parameters to implement hardware-specific control, such as spin-reversals, annealing offsets, pausing, reverse annealing, and others. There is also a separate library specifically dedicated to minor-embedding, $\texttt{minorminer}$.

An extension of the graph construction and visualization tool NetworkX~\cite{networkx}, is available through Ocean (\texttt{dwave-networkx}). This package allows users to interact directly between D-Wave software tools and the functionality of NetworkX, which is particularly useful for graph theory problems. Ocean also provides a set of classical solvers (implemented in Python) that can solve QUBO/Ising problems, such as simulated thermal annealing (\texttt{dwave-neal}) and Tabu search (\texttt{dwave-tabu}). 

In principle, any library that develops Ising/QUBO models for practical purposes can be used for quantum annealing. For example, Google's OpenFermion project~\cite{googleOpenFermion} has tools for analysis of simple quantum chemistry models using quantum computers. Although originally designed for gate-model QPUs, some algorithms (such as Hartree-Fock) can also be used for quantum annealing~\cite{streif_solving_2019}. For optimization, there is a public repository from the authors of~\cite{Chancellor2019} to implement domain-wall encoding for discrete variables\footnote{Code available here: \url{doi.org/10.15128/r27d278t029}.}.

Other commercial entities are working on developing specialized tools for quantum annealing. Of those, a promising technology is being developed by ParityQC to implement novel methods to improve quantum annealing performance~\cite{ParityQC_annealing, ParityQC_compiler}. The specific compiler being developed utilizes counter-diabatic protocols to increase the ground state probability in LHZ annealing architecture, and has the capability of addressing higher than quadratic order binary optimization problems. The ParityQC compiler also has utility for gate-model QCs, and has partnered with multiple hardware providers (both annealing and gate-model based) to continue their development.

To combat many of the issues raised in previous sections regarding solving problems directly with quantum annealers, a hybrid approach is often used. Specifically in the context of current-generation quantum annealers, hybrid algorithms are used to solve problems that are too large to embed directly on QPUs. In this approach, the original problem is partitioned into multiple sub-problems, each of which is solved using the quantum annealer, and then they are recombined to get comprehensive solution to the original problem. The first such hybrid algorithm that utilized quantum annealing processors was $\texttt{qbsolv}$~\cite{qbsolv}, which used an outer classical loop of Tabu search~\cite{glover_tabu_1997} and a D-Wave QPU in its inner loop to flip the clusters of variables found by Tabu. Later, this algorithm was open sourced, and eventually resulted in the Python Ocean library $\texttt{dwave-hybrid}$, which allows users to construct their own hybrid algorithms. The package provides multiple graph partitioning tools and classical techniques to break apart problems and merge solutions. This allows users to construct hybrid quantum-classical algorithms tailored to the specific problem being solved. It is important to note that the vast majority of industry applications with quantum annealing have used hybrid algorithms (rather than the QPU directly) because of the relatively small number of available qubits. In general, partitioning a large problem into a set of smaller problems involves difficulties and can result in sub-optimal solutions. It has been shown in practice that embedding larger sub-problems in the inner loop of the optimization that uses the QPU results in improved performance of the hybrid algorithm~\cite{Okada2019}.

In general, any algorithm that offloads some part of its calculation to a quantum processor can be considered a hybrid quantum-classical algorithm. Genetic and evolutionary algorithms have also been explored in this context, as some of the inner algorithmic steps lend themselves naturally to the QPU. Chancellor~\cite{Chancellor2017} tries to answer the question of whether QA can be used to gain advantages over modern classical algorithms, by embedding both quantum and classical search techniques, or by using multiple quantum searches sequentially.
He constructs analogues to state-of-the-art classical algorithms, but made of quantum building blocks, leading to subroutines that are less sensitive to noise in the form of problem mis-specification than QA. 
Local searches deal with small subspaces of the total solution space, and hence only errors that occur on states within these subspaces are relevant. 
A local search can therefore give meaningful results even in a problem where the global optimum is no longer correctly specified due to noise.
Chancellor \cite{Chancellor2017} suggests to use local quantum search as subroutines in well known classical algorithms like the ones from Monte Carlo's family (e.g., parallel tempering, population annealing, and isoenergetic cluster updates).

Another approach that has been proposed uses the reverse annealing feature in QA as the mutation operator in a genetic algorithm (quantum-assisted genetic algorithm, QAGA)~\cite{QAGA}.
Since reverse QA has been demonstrated for local refinement of candidate solutions, it may be considered a tunable mutation of any classically prepared state prior to the reverse annealing procedure. Here, the point $s=s'$ in the anneal ($s$ being normalized time, $0\leq s\leq 1$) determines the degree of mutation of a single candidate solution. 
Indeed, this is not possible with forward QA, since the optimization does not provide any information from the initial superposition to the final answer. 
QAGA was tested on three input classes (bimodal spin glasses, anti-cluster inputs, and deceptive cluster loops) against a suite of classical algorithms: simulated annealing (SA)~\cite{kirkpatrick_optimization_1983}, parallel tempering (PT) \cite{swendsen_replica_1986} and parallel tempering with isoenergetic cluster moves (PT-ICM) \cite{earl_parallel_2005}. 
When tested in terms of time-to-solution (TTS), QAGA performed better than QA, having a smallest variance in the output. Moreover, in comparison to classical solvers, QAGA dominated on all the input classes, even if failed to reach ground states on a limited number of inputs.

\subsection{Role of time in quantum annealing}
\label{sec:time}
There is some ambiguity in the quantum annealing community in regards to the description of time when using quantum processors (and in some cases classical algorithms). To clarify this, we define a consistent overview of the role of physical time (wall-clock time), and relate this to the QA algorithm from both a theoretical and experimental point of view. From the practical application standpoint, a QA processor represents an analog processor that samples from a semi-classical Boltzmann distribution described by a Hamiltonian~$\mathcal{H}_f$ in a predefined fixed amount of time. 
The computation proceeds in three main steps~\cite{mcgeoch_adiabatic_2014}:
\begin{enumerate}
    \item The initialization phase consists of programming the weights $h_i$ and setting the coupling strengths $J_{ij}$ of Eq.~\eqref{eq:ising_ham}. The qubits are then initialized according to $\mathcal{H}_i$. The time needed for such a programming phase is denoted $T_\mathrm{init}$.
    
    \item The annealing phase, where the transition from the initial to the final Hamiltonian according to the annealing schedule $A(t)$ and $B(t)$ from Eq.\ \eqref{eq:ising_model} is performed. 
    The physical annealing time is denoted as $T_a$, and is a user-provided parameter to the QA algorithm. In situations where reverse annealing is used, the total annealing time has two contributions, the time duration $T_\m{rev}$ of the reverse annealing plus the above mentioned forward annealing time $T_a$, i.e., $T_a^\m{tot}=T_a+T_\m{rev}$.
    
    \item The readout phase, where the qubits are measured at the end of the annealing process. The classical configuration of the spins $s_i$ is obtained, which provides a candidate solution to the optimization problem defined by $\mathcal{H}_f$. The readout time is denoted as $T_r$.
\end{enumerate}

Step (1) is performed only once per use of the QPU for a given problem $\mathcal{H}_f$. Steps (2) and (3) are repeated $k$ times to obtain proper statistics for the results and increase the probability to find the global optimum, as is typical for heuristic optimization algorithms. Therefore, the total wall-clock time to obtain $k$ samples for a given problem instance is $T_\mathrm{total} = T_\mathrm{init} + k (T_a^\m{tot} + T_r)$. Typically, the time needed to obtain the ground state\footnote{Different metrics can be used to define this time. A standard choice is time-to-solution (TTS) that is needed to observe an average probability $X$\% of obtaining the ground state, denoted by TTS$|_{X\%}$.} scales with the size of the input problem, $\mathcal{H}_f$. It is important to pause and appreciate that the scaling of total time ($T_\mathrm{total}$) relative to problem size is not the same as the ``true'' algorithmic scaling of the underlying annealing algorithm, $T_a$, and one should not be conflated with the other. The former, $T_\mathrm{total}$, is purely an engineering artifact, with constant terms $T_\mathrm{init}$ and $T_r$ that are dictated by the specific implementation of hardware, and a number of repetitions $k$ provided by the user. The latter, $T_a$, is the amount of time used for a \textit{single} annealing cycle. Theoretically, it is known that slower (longer) annealing times increase the probability of remaining in the ground state (although in practice, it is often more efficient to reduce $T_a$ and execute more annealing cycles to compensate). Thus, the smallest such $T_a$ that maximizes the ground state probability can be considered the true run-time of QA for a particular problem. 
Practically, the optimal trade-off value for $T_a$ should be long enough to give sufficiently high probability to find the ground state and at the same time should be as short as possible. The value of $T_a$ is intimately related to the problem structure and as such not known \textit{a priori}. However, $T_a$ is an input parameter to the QA heuristic, which results in a burden for the user to determine a value for $T_a$ that is appropriate for the specific problem. In this sense, the QA heuristic is not a fixed-time algorithm in the classical sense, but merely a fixed-time algorithm for one a particular problem once it has been calibrated properly. 

\section{Quantum Annealing in Real-World Applications}
\label{sec:real-world}

In this section we introduce the various domains in which there has been considerable progress related to industrial applications involving quantum annealing. We review the main works that are representative of the methods developed in those application areas, and highlight the individual contributions of each work as well as their collective importance. Due to the rapid and ongoing development in the field of quantum annealing, we note there may be papers that are relevant but not included in our summary due to scope. Ultimately, we chose the works to include based on three main criteria: original research upon which the application area is based, work that has significantly advanced each domain area, and results or methods that have considerable relevance to a practitioner. 
Table \ref{tab:extensive_review} provides a more extensive list of papers focusing on the application fields of QA discussed in this section.


\begin{table}[h!]
\centering
\caption{
Publications on real-world applications of QA. References discussed in the main text are in bold.}
\label{tab:extensive_review}
\resizebox{\textwidth}{!}{\begin{tabular}{ll}
\hline
\textbf{Application field} &
  \textbf{References} \\ \hline
Mobility &
  \begin{tabular}[c]{@{}l@{}}Traffic flow optimization:
    \textbf{  \cite{neukart_traffic_2017}, 
      \cite{quantum_shuttle}, 
      \cite{clark_towards_2019}, 
      \cite{ohzeki_control_2019}, 
      \cite{Stollenwerk2020}, 
      \cite{inoue_traffic_2021}.}
  \end{tabular} \\ \hline
\multirow{2}{*}{Scheduling and logistics} &
  \begin{tabular}[c]{@{}l@{}}Scheduling Problems:
\textbf{\cite{rieffel_case_2015},
    \cite{venturelli_quantum_2016},
    \cite{ikeda_application_2019},
    \cite{paintshop},
    \cite{yuHubbardStratQA2021},
    \cite{willsch_benchmarking_2022}}, 
    \cite{OssorioRefinerySchedulingAQ2021},
    \cite{ohzeki_breaking_2020}, 
    \cite{flight_gate_assignment}.
  \end{tabular} \\ \cline{2-2} 
 &
  \begin{tabular}[c]{@{}l@{}}Vehicle Routing Problem: 
\textbf{    \cite{feld_hybrid_2019},}
    \cite{borowski_new_2020},
    \cite{srp}.
  \end{tabular} \\ \hline
\multirow{3}{*}{Quantum Simulation} &
  \begin{tabular}[c]{@{}l@{}}Chemistry:
\textbf{    \cite{xia_electronic_2018},
    \cite{streif_solving_2019},
    \cite{mulligan_designing_2020},
    \cite{michelettiPolymerSampling2021},}
    \cite{perdomo-ortiz_finding_2012},
    \cite{hernandez_enhancing_2017}.
  \end{tabular} \\ \cline{2-2} 
 &
  \begin{tabular}[c]{@{}l@{}}Physics:
\textbf{    \cite{Harris2018},
    \cite{King2018}.}
  \end{tabular} \\ \cline{2-2} 
 &
  \begin{tabular}[c]{@{}l@{}}Biology:
\textbf{    \cite{li_quantum_2018},
    \cite{willsch_support_2020}.}
  \end{tabular} \\ \hline
\multirow{4}{*}{Machine Learning} &
  \begin{tabular}[c]{@{}l@{}}Classification: 
\textbf{    \cite{mott_HiggsOpt_QA_2017},\cite{nevenBinaryClassifierAnnealing2009}, \cite{nevenBinaryClassifierAnnealing2009_2}}.
  \end{tabular} \\ \cline{2-2} 
 &
  \begin{tabular}[c]{@{}l@{}}Reinforcement learning:
\textbf{    \cite{neukart_quantum-enhanced_2018},}
\cite{crawford_reinforcement_2019}.
  \end{tabular} \\ \cline{2-2} 
 &
  \begin{tabular}[c]{@{}l@{}}Cluster analysis:
\textbf{    \cite{quantum_clustering},
    \cite{kumarClustering2018}.}
  \end{tabular} \\ \cline{2-2} 
 &
  \begin{tabular}[c]{@{}l@{}}Matrix factorization:
\textbf{    \cite{nbmf},
    \cite{nbmf_ra1},
    \cite{nbmf_ra2}.}
  \end{tabular} \\ \hline
Finance &
  \begin{tabular}[c]{@{}l@{}}Portfolio optimization: 
\textbf{    \cite{optimal_trading},
    \cite{elsokkary2017financial},
    \cite{Venturelli2019},
    \cite{MugelPortfolio2020},
    \cite{grant_benchmarking_2021},}
    \cite{rosenberg_solving_2016},
    \cite{phillipson_portfolio_2020}.
  \end{tabular} \\ \hline
\multirow{2}{*}{Miscellaneous} &
  \begin{tabular}[c]{@{}l@{}}Finite-element design: 
      \cite{VreumingenFiniteElements2019}.
  \end{tabular} \\ \cline{2-2}
  &
  \begin{tabular}[c]{@{}l@{}} Material design:
   \cite{kitaiQA_FA_2020}, \cite{wilsonbVAEEngOpt2021}.
  \end{tabular} \\ \hline
\end{tabular}}
\end{table}

\subsection{Traffic flow optimization}
\label{subsec:tfo}
The problem of traffic congestion on a road network occurs if the demand volume exceeds the road capacity, leading to important problems such as delays, increased fuel consumption, and additional pollution. 
Optimizing traffic flow hence represents one of the real-world challenges that can be modeled by using different strategies, such as classical optimization (see Refs.~\cite{armasTrafficOpt2017,boetherTraffic2021,trafficOpt2018} for recent examples), reinforcement learning \cite{rezaee_application_2012,selvaraj_multi-agent_2015,walraven_traffic_2016} and neural networks \cite{kwon_comparative_1994,wei_fl-fn_2002}.
More recently, solutions that take advantage of QA technologies have been also proposed. 

Neukart et al.~\cite{neukart_traffic_2017} were the first to show how to transform  real-world data -- GPS coordinates in this particular case -- and formulate a relatively simple QUBO to be solved as a real-time application.
Their target was to minimize the congestion in road segments on overlapping routes in order to minimize the time for a set of cars to travel between individual sources and destinations.
The dataset tested was composed of 418 cars with 3 alternative routes generated for each car to redistribute traffic, which resulted in 1254 logical variables to represent the problem.
Using the hybrid-quantum solver \verb|qbsolv|, they partitioned the problem into sub-problems targeted for execution on a D-Wave 2X system, obtaining a significant improvement over randomly uniform assignment of cars to routes.
Hence, although a simplified version of traffic flow (limited set of cars, no communication to infrastructure, no time window) was presented, the paper showed that some types of real-time optimization problems can be posed for the D-Wave systems and the hybrid tools that use them.

Theory was put into practice with a follow-up by Yarkoni et al.~\cite{quantum_shuttle}. In this paper, the authors implemented the paradigm proposed in~\cite{neukart_traffic_2017} to navigate a small bus fleet in real time for the duration of the Web Summit 2019 conference in Lisbon, Portugal. Volkswagen AG partnered with the local city authorities of Lisbon to present the project, named the “Quantum Shuttle”, which operated a public shuttle bus service between the Web Summit conference venue and the city center of Lisbon.
To adapt to a real-world scenario, GPS points of live traffic data are used in the objective function definition instead of road segments.
The largest QUBO that was solved was quite limited in size (12 variables, with 5 buses being navigated concurrently), and three different methods of using QA were tested: direct embedding, hybrid algorithms, and D-Wave's Hybrid Solver Service.
Each of these methods was integrated into a custom-build quantum optimization service build using cloud computing which was used to navigate the bus fleet. Thus, this was the first commercial application that depended on a quantum processor to perform a critical live task.

Another extension of~\cite{neukart_traffic_2017}, by Clark et al.~\cite{clark_towards_2019}, showed how to route multiple robots on a grid in real time. While their approach is similar to the original, the authors presented an additional constraint to avoid collisions (or overlaps) in time.
Their objective was to find a route of minimum cost for each robot, where the cost is dependent on the distance the robot travels and the change in direction in the chosen route.
They also used the decomposing solver qbsolv to propose a hybrid algorithm that was able to address a maximum problem size of 800 variables. 
In terms of time to solution, the performance of QA was promising, as qbsolv with the D-Wave 2000Q quantum annealer scales significantly better than with a classical solver.
However, performance bottlenecks were evident on both the classical part (creating the QUBO problem and collision detection) and the quantum part (communication and embedding time) of the hybrid algorithm.
To reduce the execution time, they argued that the size/number of subQUBOs sent to the quantum annealer needs to be maximized/minimized, resulting in fewer calls to the QPU and lower communication and embedding times. This highlights a high dependence of the solver on the connectivity of the graph, and overall performance can be improved with more connected hardware architectures.

Further work based on~\cite{neukart_traffic_2017} was proposed by \citet{ohzeki_control_2019}. The novelty is that they considered the time dependence on the binary objective characterizing a certain route.
The application in this work was the controlling problem of automated guided vehicles (AGVs), which are portable robots for moving materials in manufacturing facilities and warehouses.
The authors considered problem instances with 10 AGV and 6 candidate routes leading to a maximum size of 60 variables, which can be directly solved on the D-Wave 2000Q hardware. Lager problem instance up to 900 variables are solved using \verb|qbsolv|.
In this case, comparison of the QA performance to alternative approaches using the Fujitsu digital annealer~\cite{aramon_QA_DA_2019} and \verb|Gurobi| software solver\cite{gurobi} did not show a clear benefit of utilizing the QA hardware. 
However, the authors estimated the performance of the quantum annealer by measuring the actual computational time, which includes communication and queuing time in cloud services. If machine time is instead considered, the D-Wave 2000Q can be a good candidate in this industrial application.

\citet{Stollenwerk2020} solved the conflict resolution problem for wind-optimal airplane flight trajectories. The trajectories are obtained by minimizing the fuel cost over the routes with given origins and destinations and desired departure times under consideration of wind forecasts.
Here too, the presented version of the problem is a simplification of the real-world use-case since they do not consider all the physically realistic trajectories, but rather perturb trajectories by delaying the corresponding flight on the ground prior to departure.
The heuristic embedding algorithm described in \cite{embedding} was used to embed instances with up to 50 flights and 104 conflicts on the D-Wave 2X, as well as up to 64 flights and 261 conflicts on the D-Wave 2000Q.
The paper also introduces a novel pre-processing algorithm to eliminate potential conflicts that, given a maximum delay, can never occur and clustering adjacent conflicts. The authors concluded that, for reasonable modeling choices, the most challenging sub-problems that are programmable in the current devices are solved to optimality with 99\%  probability within a second of annealing time. However finer model discretizations, as well as larger problem sizes, decrease the success probability due to the limited precision QA processors.

Very recently, Toyota Central R$\&$D Labs also presented a study aimed to globally control traffic using signals arranged on a square lattice with a D-Wave 2000Q quantum annealer \cite{inoue_traffic_2021}.
They follow a 2-step approach. First, they formulate a signal optimization problem that minimizes the imbalance of traffic flows in two orthogonal directions. Afterward, they reformulate the problem as an Ising Hamiltonian. 
The Hamiltonian is divided into 42 groups (by the graph partitioning technique in the Ocean library~\cite{dwave_ocean} provided by D-Wave) and the optimization problem is solved in parallel.
For comparison purposes, the problem was also solved using SA, but QA performed better in the specific parameter domain. 
Moreover, the difference between the time average of the cost function with problem partitioning to a problem without partitioning seemed to increase with the rate at which vehicles go straight at intersections, which indicates a performance drawback of the partitioned problems.

In conclusion, all presented methodologies to address traffic flow optimization problems are based on assumptions that make them far from faithfully representing real-world problem challenges. 
However, QA processors have been proven to be valuable tools to address simplified formulations and show high potential, in particular if we take into account the rapid development of successive generations of QA systems over the recent times.

\subsection{Scheduling and logistics}
\label{subsec:sched_log}
Many planning applications include scheduling, which deals with allocating resources and time to tasks while taking constraints into account \citep{chien_generalized_2012}. 
While many studies were still looking at the performance of quantum annealers on problems native to their architecture, \citet{rieffel_case_2015} explored general mappings to QUBOs for hard combinatorial optimization problems that arise in operational planning, and applied them to two parameterized families of planning problems: navigation-type planning problems and scheduling-type planning problems.
In particular, they explored two different approaches for mapping problems to QUBOs, which are the time-slice and conjunctive normal form (CNF) methods. They provided comparisons of the effectiveness of different embedding mappings onto hardware and parameter choices for their solution on the 509-qubit D-Wave Two machine.
In all cases, they used an annealing time of 20 $\mu s$ to take full advantage of the machine capabilities (the shortest times available).
They considered a range of problem sizes from 4 to 12 for the navigation-type and 8 to 16 for the scheduling-type problems.
Although their study did not produce results competitive with state-of-the-art classical approaches, they investigated  the embeddability in future architectures and stated that increasing the local connectivity within QPU unit cells would have a much greater impact on the ability to run instances of applications than simply increasing the number of unit cells.

Later, \citet{venturelli_quantum_2016} presented in detail an implementation of the job-shop scheduling problem (JSP) that makes
use of the same D-Wave quantum annealer.
They formulated a QUBO version of the JSP decision problem as a constraint satisfaction problem (CSP): optimizing the allocation of resources in the execution of sequences of operations was done by encoding constraints on location and time.
The JPS involves a set of jobs (sequence of operations) that must be scheduled on a set of machines, with the final target of minimizing the makespan, i.e., the completion time after the run of the last job is over \citep{janiak_solution_2009}.
The analysis was performed on up to $N=M=9$ and $p_\text{max}=20$ using $200$ instances, where $N$ is the number of jobs, $M$ is the number of available machines, and $p_\text{max}=20$ is the maximum execution time of the operations.
They compared the performance of the D-Wave QPU to two classical algorithms: the branch and bound algorithm \citep{brucker_branch_1994} and the MS algorithm by \citet{martin_new_1996}. While the QPU performed better than the MS method, the QPU took longer than the branch and bound algorithm to generate valid solutions.

More recently, \citet{ikeda_application_2019} addressed the Nurse Scheduling Problem (NSP) with hard constraints using the D-Wave 2000Q quantum annealing device.
They formulated the problem as a constrained QUBO and implemented it through the randomized embedding algorithm based on~\citet{cai_practical_2014}.
With $N$ as the number of nurses whose shifts are to be scheduled on $D$ working days, then the optimization problem consists of $N \times D$ variables, where the single variable specify the assignment of nurse $n$ to day $d$, for $n = 1, \dots, N$ and $d = 1, \dots, D$.
The authors compared QA to simulated annealing for $N=\{3,4\}$ and $D = \{5,6,\dots,14\}$ and drew statistics on 1000 samples for 10 shots for D-Wave 2000Q and 500 annealing times 10 shots for SA.
The authors tested both forward and reverse annealing and showed that for a fixed sample size and fixed annealing duration, the success probability of completely satisfying all constraints decreases as the size of the schedule $D$ and the size of the roster $N$ increase. Although reverse annealing improved the success probability, this did not happen uniformly, and simulated annealing gave better results, highlighting the need for further investigation into the parameters of the annealing process.

Also related to JSP is the vehicle routing problem (VRP)~\citep{toth_vehicle_2002}, which is a combinatorial optimization and integer programming problem to determine the optimal set of routes to be taken by vehicles to serve a given set of customers as efficiently as possible.
Belonging to the category of NP-hard combinatorial problems, \citet{feld_hybrid_2019} were the first to address the capacitated vehicle routing problem (CVRP), where each vehicle has a capacity constraint on how much it can transport, by mapping it onto a QUBO formulation and solving it using a D-Wave’s 2000Q quantum annealing hardware.
The problem sizes considered in this work ranged from 14 to 38.
In order to fit the QUBO directly onto the QPU, they used the minor embedding technique by \citet{cai_practical_2014} in combination with D-Wave's \verb|qbsolv| tool, leading to a sub-problem size of 20 logical variables.
In particular, they presented a hybrid method based on a two-phase approach (a clustering and a routing phase) which could compete with other classical and 2-phase heuristics, and in some cases even provided better solution quality.
However, this hybrid method did not provide overall advantages in terms of solution quality or computation time compared to classical heuristics, which, according to the authors, was due to both the size of the hardware used and the need to use additional tools such as \verb|qbsolv|.


The problem of scheduling the charging processes of a fleet of electric vehicles (EVs) was addressed by~\citet{yuHubbardStratQA2021}.
The task was to charge  $N_\m{bus}$ EV buses on $N_\m{station}$ charging stations where charging is only possible in predefined time windows distributed over the day.
The objective was to determine the charging schedule for a bus fleet for a complete day which minimizes the total electricity cost, where the electricity prices were given as part of the problem specification.
Additionally, the charging process for each bus needed to respect a constraint on the allowed state of charge of the battery to be within 30\% and 100\% capacity. 
The authors study various problem instances with varying numbers of buses and charging stations where the maximal problem size consisted of about 220 logical variables for an instance with three buses and two charging stations. In order to incorporate the problem constraints into the QUBO, the authors employed the method based on the Hubbard-Stratonovich transform described in Sec.~\ref{sec:constrained_opt}. With this, the problem could be directly embedded on the D-Wave 2000Q hardware without partitioning. Even though the authors could show that they were able to consistently find valid solutions with the QA, the comparison to SA showed that it took more iterations to arrive at a solution and that the solution quality found with QA was not better than those of SA.

A factory scheduling problem was presented by~\citet{paintshop}, derived from a real-world optimization problem involving re-ordering color sequences in a paint shop during the manufacturing of cars. The task was to assign colors to sequences of cars such that all customer orders are fulfilled while simultaneously minimizing the number of color switches within the sequence. In this work, the authors investigated problem instances ranging from 10-3000 cars in a single sequence. The major contribution of this work is that the multi-car paint shop problem described in the paper has a simple Ising model representation, and instances derived directly from real-world data were solved using D-Wave QPUs, hybrid algorithms, and classical algorithms. The results showed that while the quantum hardware and hybrid algorithms were able to provide adequate solutions to the problem for smaller sizes, the performance of these algorithms approached that of a simple greedy algorithm in the large size limit. However, the analysis also concluded that QA is approaching the limit of industrially-relevant problem scales.   

\citet{willsch_benchmarking_2022} recently compared the latest-generation D-Wave Advantage and the D-Wave 2000Q annealers with small and large-scale exact cover problems representing simplified aircraft scheduling scenarios.
As expected, they showed that a considerable increase in performance can be reached on Advantage annealers, in terms of success rate, speed for programming and readout, and time to solution. 
In addition, smaller fluctuations in success rates were observed.
However, increasing the number of qubits and couplers in the annealer does not necessarily improve performance independently of the  considered problem instance. The authors could obtain better performance on D-Wave 2000Q for problems with very sparse connectivity for which more efficient embedding is not needed, meaning that additional couplers only introduce additional noise when the problem instance does not require them. Thus, the superiority of Advantage over D-Wave 2000Q lies in the ability to use better embeddings, if necessary, rather than in the improved architecture itself. Although the D-Wave Advantage QPU is overall an improvement relative to its predecessors, the choice of the quantum annealer must be based on the characteristics of the problem being studied.

Currently, the scheduling-type problems that can be solved on current and near-term QA hardware are already somewhat realistic and not too far away from real-world applications. The special structure of those problems, especially the constraints typically employed, can lead to small problems instances being difficult and even infeasible to solve with classical approaches already (for example Refs.~\cite{JainJSSReview1999, kuMIP_JSS2016,IshiharaEVCharging2020}). Such small but hard problem instances may be particularly interesting for quantum approaches, and consequently, scheduling problems seem to be a very promising field for future application of QA to real-world problems.

\subsection{Quantum Simulation for Chemistry, Physics and Biology}
\label{subsec:quantum_chemistry}
The original motivation of quantum computing was the realization of a programmable quantum system that is capable of simulating general quantum systems. 
To simulate the full dynamics of complex molecules, prototyping efficient batteries, simulating photosynthesis processes in organic materials, and other challenging problems, finding the ground state of individual atoms and molecules is a problem characterized by an extremely high complexity due to exponentially growing Hilbert spaces.
QA processors, being complicated quantum mechanical systems themselves, can represent an option to find solutions to these complex problems.

\citet{streif_solving_2019} followed an approach outlined by \citet{xia_electronic_2018} to map electronic structure Hamiltonian calculations to a classical spin-glass system, and subsequently obtain a precise estimation of the true ground state energy of the molecule with QA.
In their research, they computed the ground state energies of molecular hydrogen (H$_2$) and lithium hydrid (LiH) for various interatomic distances on a D-Wave 2000Q system.
To accomplish this, since a direct embedding of the terms containing Pauli matrices in the electronic structure Hamiltonian is not possible, they introduced $r$ ancillary qubits for each of the $n$ qubits of the original Hamiltonian. 
The factor $r$ is called \textit{scaling factor}.
Although the investigated method works in general since it is able to provide energy estimations very close to the exact energies, due to the currently limited Hamiltonian terms in the QPU, in general the transformations in~\cite{xia_electronic_2018} incur exponential overhead in transforming the quantum simulation to a spin-glass. The authors underlined that larger systems that were difficult to address might become accessible with more efficient embedding techniques and reverse QA.

\citet{mulligan_designing_2020} address the protein design problem  by using a D-Wave 2000Q system.
They expressed the optimization problem as a QUBO in such a way that the QPU could interface with Rosetta, a state-of-the-art design software suite that used to address protein design tasks by using simulated annealing-based heuristics. 
The quantum design algorithm was compared with a classical exact branch-and-bound algorithm and the classical simulated annealing method used by default by Rosetta.
Results were based on 400 design tasks with the number of designable positions ranging from 1 to 40 and the geometric number of rotamers per position ranging from 1.03 to 6.77, so that the tasks could be implemented in the QPU. 
Additional 5376 higher-dimensional tasks were addressed with the classical strategies, which could not be addressed with the QA-based strategy. 
While the classical branch-and-bound and simulated annealing strategies scale exponentially with the problem size (as expected), the QA-based strategy exhibits constant time-to-solution. However, for the small problem instances addressable on the D-Wave 2000Q system, the absolute time to arrive at a solution with the QA-based approach was found to be one to four orders of magnitude slower than the classical approaches.

The problem of sampling valid ensemble configurations of dense polymers mixtures was addressed by \citet{michelettiPolymerSampling2021}. 
The problem was formulated as sampling self-avoiding walks and rings on a predefined lattice for dilute to dense polymer concentrations. The QUBO formulation incorporated terms for enforcing the number of monomers and bonds,  the chain connectivity as well as self-avoidance, enabling direct control over the types of polymers to consider. The authors empirically observed an exponential decrease in the success rate of finding valid ensemble samples with increasing problem size for classical simulated annealing. In contrast, the hybrid quantum-classical QA approach tested did not suffer from the same deficiencies.

\citet{Harris2018} used a D-Wave quantum annealer to simulate the phases of a correlated electron system and investigate the properties of a quantum magnet modelled by the transverse-field Ising Hamiltonian. In general, these systems are difficult to simulate with limited computational resources.
The authors tuned the amount of frustration in a three-dimensional cubic lattice of dimensions $L \times L \times L$ and varied the transverse field in order to observe phase transitions between a paramagnetic, an antiferromagnetic, and a spin-glass phase.
Experiments were conducted for three system sizes $N = L^3$, with $L = \{4, 6, 8\}$, and results were based on ensembles of up to $1000$ random instances, where each instance corresponded to specific sets of ferromagnetic/antiferromagnetic interactions between the spins.
The results compared well to the theoretic results and thus validated that the transverse-field Ising model can be used in an integrated circuit as a materials physics simulator, thanks to the ability to manipulate individual spins and bonds.

\citet{King2018} also simulated a transverse-field Ising model to investigate the Kosterlitz-Thouless (KT) phase transition, a specific type of topological phenomenon on frustrated lattices.
In particular, the authors used a D-Wave QPU to simulate a fully-frustrated  square-octagonal and triangular lattice with cylindrical boundary conditions of size $L \times L$.
These lattices were embedded onto the Chimera connectivity graph of a D-Wave's 2000Q processor.
Since the translation of the geometry of the lattice onto the Chimera topology leaves necessarily unused qubits along the outside boundary of the processor, the largest instance studied, for $L=15$, used $1800$ of the $2048$ qubits available in the processor.
By using a reverse annealing protocol, the authors could successfully realize the exotic KT phase of matter. 
The flux qubit implementation on the D-Wave hardware agreed with theoretical predictions and quantum Monte Carlo simulation over a range of system sizes and Hamiltonian parameters, thus highlighting the potential of QA processors to simulate systems of which classical simulations are hard or even intractable. 

Biology is another field characterized by high-complexity dynamics. 
The first application of QA to real biological data was proposed by \citet{li_quantum_2018}, who explored the possibility of using D-Wave Systems QPU to solve the simplified formulation of a biologically relevant transcription factor (TF)-DNA binding problem.
Given a transformed feature vector that represents a DNA sequence, their goal was to compute a predicted binding score that best matches the actual binding score.
Although the D-Wave 2X processor used for this study had 1098 functional qubits, because of a sparse connectivity between qubits, only about 40 features could actually be implemented on the device.
For small training sizes, the quantum annealer performed better than its classical counterparts (simulated annealing, simulated quantum annealing, multiple linear regression, LASSO, and extreme gradient boosting) for classification, and competitively for ranking tasks. However, the authors conclude that for larger datasets classical algorithms may be a better choice.

In Ref. \citep{willsch_support_2020}, the authors studied the implementation of kernel-based support vector machines (SVMs) on a D-Wave 2000Q quantum annealer, by representing the objective function of the SVM as a QUBO.
They applied both the classical SVM and their version of SVM using QA to synthetic two-dimensional data and real data from biology experiments.
To measure the classification performance, they used an experimental dataset with 4352 training samples and 454 test samples. The results showed that the ensemble of classifiers produced by the quantum annealer was superior in terms of classification accuracy to the single classifier obtained by the classical SVM for the same computational problem. Therefore, the quantum version of SVM has potential in practical application for classification problems in computational biology, especially when the dataset is rather small.

\subsection{Quantum Machine Learning}
\label{subsec:quantumML}

Quantum machine learning refers to quantum algorithms that are used to solve machine learning tasks, for
instance, the outcome of the measurement of a qubit can represent the result of a binary classification task.
Here we discuss different machine learning problems that have been addressed by making use of QA: classification~\cite{mott_HiggsOpt_QA_2017} and reinforcement learning problems~\cite{neukart_quantum-enhanced_2018}, cluster analysis \cite{quantum_clustering, kumarClustering2018}, and matrix factorization~\cite{nbmf,nbmf_ra1,nbmf_ra2}. 
For an introduction on the general topic, see, for example, Refs. \cite{mishraQMLReview2021, nath_review_2021}.

Early applications of QA to classification problems can be found in Refs.~\cite{nevenBinaryClassifierAnnealing2009,nevenBinaryClassifierAnnealing2009_2}, where the authors propose a boosting-based algorithm and compare it to the AdaBoost approach on simple benchmark examples. The authors conclude that the proposed QA based approach shows advantages over AdaBoost as the generalization error is lower and the classification is faster as it involved less weak learners and fewer boosting steps.

Mott et al.~\cite{mott_HiggsOpt_QA_2017} implemented a classifier on classical and quantum annealing QPUs to solve a Higgs particle signal-versus-background machine learning optimization problem. 
The authors built a set of weak classifiers based on the kinematic observables of the Higgs boson signature through photon decay, which was then combined to construct a strong classifier.
Implemented on the D-Wave 2X system, the optimization involved 36 logical binary variables used for the weak classifiers, and 432 physical qubits.
QA and SA were compared to traditional machine learning methods -- a deep neural network and an ensemble of boosted decision trees -- using the area under the receiver operating characteristic curves as performance metric for different sizes of the training dataset. Results showed that for very small sizes of the training dataset the quantum annealer-trained classifiers had  some advantage over traditional tree- and deep neural network-based machine learning methods, but still comparable to simulated annealing. For larger training sets, the performance of simulated as well as quantum annealing was inferior to traditional methods.

An example of QA in the area of reinforcement learning was provided by \citet{neukart_quantum-enhanced_2018}, who showed how to partially embed policy evaluation for discrete state spaces on a D-Wave QPU, and to find quantum-classical algorithms capable of dealing with continuous state spaces.
By taking the card game blackjack as an example, their aim was to approximate the optimal policy under consideration of all states from a certain number of complete episodes.
From 2 to 7 sub-optimal state-value functions were generated, using 100-500 episodes each.
It was shown that quantum-enhanced Monte Carlo policy evaluation allows for finding equivalent or better state-value functions for a given policy with the same number episodes compared to a purely classical Monte Carlo algorithm.
However, the authors make clear that the target of their study is to explain how to represent and solve parts of the concerned problems by using QPU, and not to prove an advantage over every existing classical policy evaluation algorithm.

A quantum-assisted clustering algorithm (QACA) was introduced by \citet{quantum_clustering}. 
The algorithm is described as showing similarities with the self-organizing feature map, in the sense that the topological properties of the D-Wave QPU are exploited for cluster assignments. 
Moreover, the algorithm is developed in such a way that a singular embedding, without splitting the QUBO, is possible.
The Iris data set (150 points) was used for verification and QACA was compared to expectation minimization, $k$-means, and self-organizing feature map approaches.
Each data sample had three features represented by three logical qubits, which lead  to a total of 450 logical qubits.
It was shown that QACA can compete with the classical algorithms in terms of accuracy. However, the quality of the results strongly depend on the cluster-form, insinuating that improved techniques for cluster-form initialization must be developed. The authors conclude that the main advantage of QACA is that it does not require training in contrast to classical algorithms, but the main disadvantage is given by necessary embedding onto the QPU. 
A similar approach was taken by~\citet{kumarClustering2018}, which  compared $k$-means clustering with one-hot encoding and binary clustering for problem instances with up to 2000 data samples and two to six clusters to be clustered. 

D-Wave quantum annealers have been also used to perform matrix factorization. 
\citet{nbmf} focused on learning features in a set of facial images based on non-negative binary matrix factorizations (NBMF), and developed an algorithm implemented on a D-Wave 2X QPU. 
Given a matrix $V$, the NBMF looks for a decomposition of the matrix $V=WH$ such that $W$ is composed of non-negative floating point numbers and $H$ is binary.  
To do so, the authors used an alternating least-squares algorithm \cite{lin_projected_2007} where QA is used for finding the matrix $H$ in such a way that, given a matrix $V$, the norm between $V$ and the matrix product $W \cdot H$ is minimized. 
Although $H$ can be found by solving a series of linear least squares problems in binary variables, the same problem can be readily solved as a QUBO.
The analysis was limited to 35 binary variables, each represented by at most 19 physical qubits, leading to a total of 660 physical qubits in the worst case scenario.
In this particular study, QA outperforms two classical approaches (\verb|qbsolv| and \verb|Gurobi|\cite{gurobi}) in terms of TTS when low to medium number of annealing cycles were sufficient to find a NBMF on a benchmark that was not tailored to the QPU as other problems have been.
Moreover, although QA does not outperform the best classical heuristics in general, it can be considered the best choice in case only a short computational time is available.

The same authors \cite{nbmf_ra2} as well as other researchers \cite{nbmf_ra1} addressed matrix factorization with the D-Wave 2000Q system by exploiting its reverse annealing feature, which
performs series of local searches that refine existing solutions obtained through forward annealing.
\citet{nbmf_ra2} presented an algorithm to overcome the downside of the QA approach, which was that improvement in solution quality from iteration to iteration quickly plateaus due to the global search nature of forward annealing search.
It was observed that, in terms of the quality of the factorization given a fixed time, reverse annealing performs well once the error of the factorization is set sufficient low, i.e., lower than the plateau reached with forward annealing, leading to an average improvement on the quality of the NBMF factorization by 12\%.
However, the exact nature of the matrix being factorized appeared to play a role in determining how effective the algorithm is, which highlights the need for further investigation on the topic.

\citet{nbmf_ra1} went one step further, by addressing the same decomposition problem as in~\cite{nbmf}, but where neither $W$ nor $H$ are binary matrices. 
The ultimate goal was to make it possible to use QA in real-world applications that do not directly involve binary variables.
They also introduced an adaptive strategy to maximize the efficiency of the reverse annealing. The algorithm searches for the right value of the mean of Hamming distances by adapting the holding time parameter of the reverse annealing search. 
Results showed that the proposed method was able to find the global minimum point and required much less iterations to reach good factorizations compared to classical algorithms.
Although a very limited number of variables was considered, this study showed that possible applications for QA could include more variable types than strictly binary.

So far, no clear advantage of QA-based approaches over classical machine learning could be demonstrated, but some aspects are promising. However, many of the favorable features of QA approaches only manifest themselves if specific conditions are met, for example, sparsity of the data or small data set sizes.

\subsection{Finance}
\label{subsec:finance}

Many applications of quantum computing and in particular QA are proposed for the finance domain \cite{orusFinancialQCReview2019}, but approaches that actually use real QA hardware are rare. 

Portfolio optimization represents a real-world application area in finance which one could use to benchmark QA.
In particular, QA has been applied to a well-deﬁned class of unconstrained
optimization problems derived from the application of Markowitz portfolio theory \cite{markowitz_portfolio_1952}. 
Markowitz portfolio optimization is a variant of binary optimization tasked with selecting the subset of investment assets expected to yield the highest
return value and minimal risk and transaction costs, including temporary and permanent market impact.
Several versions of portfolio optimization problems have been used to benchmark QA performance \cite{optimal_trading,elsokkary2017financial,Venturelli2019,grant_benchmarking_2021}.

\citet{optimal_trading} demonstrated several encodings of a discrete multi-period Markowitz portfolio optimization formulation using D-Wave Systems QPUs, and found promising initial results for the probability to find the optimal portfolio.
Mean-variance portfolio optimization problems are traditionally solved as continuous-variable problems, while in this case a discrete solution is necessary.
The portfolio optimization problem was written as a quadratic integer optimization problem, where integer variables were recast as binary variables, and the constraints were incorporated into the objective function as a penalty term.
The study presents a double encoding: the first one from integer to binary variables (binary/unary encoding), the second and standard one from the logical graph to the physical graph, using 584 physical qubits to represent up to logical 72 variables.
The analysis highlighted a level of intrinsic noise that manifests as a misspecification of the optimization problem. The authors conclude more qubits would allow the use of an integer encoding scheme that is less sensitive to noise levels.

\citet{elsokkary2017financial} also tested whether the use of a D-Wave One QPU allowed for a meaningful improvement in computational performance for solving the Markowitz portfolio for data gathered from the Abu Dhabi Stock Exchange website in Abu Dhabi, United Arab Emirates.
They formulated -- with some relaxations since fractional shares of stocks cannot be purchased -- the Markowitz portfolio model as a QUBO and compared the solutions found with QA to the ones obtained through a MATLAB's genetic algorithm solver.
The data set was too large for direct embedding, but \verb|qbsolv| showed good agreement with the MATLAB-derived solutions.

\citet{Venturelli2019} benchmarked a similar mean-variance model of portfolio optimization using a hybrid quantum-classical method that coupled quantum annealing with a genetic algorithm (GA). 
The authors found that the best approach was to seed a D-Wave 2000Q quantum annealer with a candidate solution found by a greedy local search and then perform a reverse annealing protocol.
Since all the instances were also solvable by an industry-established genetic algorithm, they used the latter as the classical benchmark.
From the results comparing GA, QA and reverse QA, the best solver was reverse QA at minimum annealing time and pause time.
In the median case, one to three orders of magnitude speed-up was observed when applying the proposed hybrid algorithm with respect to forward annealing alone.

\citet{MugelPortfolio2020} performed portfolio optimization using the D-Wave Advantage QPU, which features a qubit connectivity based on the Pegasus topology, i.e., 5000 qubits and 15-way qubit connectivity.
The authors also showcased the performance of a D-Wave proprietary hybrid algorithm available in the Leap quantum cloud service, called the Hybrid Solver Service (HSS), which had been used in previous studies as well~\cite{quantum_shuttle}.
\citet{MugelPortfolio2020} solved the portfolio optimization problem for two indices, the Nikkei225 and the S\&P500, and compared the results of HSS with simulated annealing, a genetic algorithm, the \verb|Gurobi| solver~\cite{gurobi}, and a \verb|LocalSolver|~\cite{LocalSolver}, which combines exact and heuristic techniques.
Among these solvers, \verb|LocalSolver| performed the best and found the optimal solution very quickly in all cases.
In terms of TTS, HSS was able to find a solution within 3\% of the optimal solution for the S\&P500 index up to 400 stocks.

More recently, \citet{grant_benchmarking_2021} presented a study to benchmark QA controls and their relative effects on computational accuracy using portfolio optimization. 
They evaluated forward and reverse annealing methods and identified control variations that yield optimal performance in terms of probability of success and probability of chain breaks.
They focused on the controls during the phases of pre-processing (Hamiltonian definition and embedding onto the quantum annealer), annealing (control on the time-dependent physics of the device and underlying quantum state), and post-processing (readout and decoding of observed results).
In particular, a variety of quantum annealing controls were explored: the embedding algorithm, the forward annealing time, the number of spin reversal transforms, reverse annealing initial state, annealing pause point and
duration, and others.
The novelty of this paper is that the authors provided insights into the best selections of controls for solving binary optimization problems with the highest accuracy that may help guide the future use of QA as a metaheuristics for optimization of industrially relevant applications.

In summary, it is not clear whether QA will be the most compelling solver for finance applications, but as technology and theory advance, it could be a competitive choice.
These results also highlight the importance of hybrid quantum/classical approaches that allow for solving much larger problems than those that can be embedded in the topology of current quantum processors.

\subsection{Final remarks on QA for real-world applications}
\label{sec:final_remarks_applications}
In summary, quantum annealing methods have been tested for numerous applications, from traffic flow optimization and job-shop scheduling to machine learning, portfolio optimization, protein folding, molecular similarity, and computational biology.
The various optimization problems presented here share the solution strategy of formulating QUBO problems, which imply freely programmable and typically fully connected interactions between Ising variables. The mapping from these dense logical graphs to the sparse connectivity in QPUs prevents efficient computation in the current generation quantum annealers.
This is probably the most crucial and common bottleneck of QA for real-world applications.

What we can learn from current applications of QA to traffic flow optimization is that all problems are based on two main models: either every variable is represented by a node of a graphical network and optimization is directly performed on the network itself, or decision variables represent properties of the graphical network and constraints are imposed on these logical variables. All traffic flow optimization problems consider paths and congestion. However, time structure is often missing, which is crucial if one wants to consider real-world application cases. 
On the other hand, we can state that scheduling applications are closer to real-world needs. Problems that are not far from some problems found in practical life are already addressed, even though still limited by the available hardware size.
A similar problem arises in financial applications. Indeed, what to do in real financial problems is beyond the capabilities of quantum annealers. Moreover, the time required to execute these algorithms is much longer than that required by financial traders to be considered useful. Due to this timing delay, it is impossible to solve real-world finance problems directly using QPUs, and some mathematical modeling is therefore required. This in turn opens the possibility of using classical competition algorithms, making it difficult to argue how QA can offer an advantage.

Conversely, in the field of chemistry, simulation, or ML, the timescales of QA offer an advantage rather than a bottleneck. Sampling from complex crystal structure ground states or mathematical models useful for machine learning algorithms is hard, and long timescales are typical classically. However, the scaling of the relevant terms for impactful applications in these fields is disadvantageous to QPUs. For ML, it is standard to have many layers of densely connected variables, something that is directly orthogonal to the regular sparse topology of current quantum devices. While there is more potential for quantum simulation, as evidenced by recent works in which D-Wave QPUs successfully simulate previously intractable quantum crystal models, the interaction terms and limited size of the QPUs offer strict limitations in the kinds of simulations that QPUs can execute. Furthermore, decomposing arbitrary quantum simulations into simpler models incurs an exponential overhead in the number of qubits, leading to unfavorable scaling of the problem size.

While it is clear that realistic applications using QPUs are still far, it is important to understand the different limiting factors. Most obviously, the number of qubits and couplers in current QPUs is low relative to the number of variables typically used in each of the application domains. An ideal practical problem needs to be small enough for significant parts of it to fit on current hardware, but also large and hard enough that it cannot be solved quickly and efficiently by a purely classical solver. Moreover, QA is a heuristic algorithm, which means that a QPU is able to provide approximations of optimal solutions with each annealing cycle, but may require a much larger number of anneals to find a global optimum. In the worst case, it is not even always guaranteed to get feasible solutions at all. Finally, QA benefits from a classical support framework that provides tools for transforming, decomposing, and embedding inputs. However, there is still much room for improvement, both in terms of performance and scope. In the near future, it is hoped that hybrid methods will outperform pure classical or quantum methods in some -- application-specific -- cases.
Together with improvements in QPU size and performance and more efficient infrastructure support, the development of efficient and effective hybrid algorithms will hopefully combine the best features of classical and quantum processors: finding optimal solutions and doing so quickly.

\section{Lessons Learned about Quantum Annealing}
\label{sec:lessons}

The main goal of this work is to provide researchers and practitioners from various fields with a condensed source explaining the principles of QA, as well as summarizing the shortcomings and potential benefits of QA in view of real-world application problems. 
Considering the body of work presented and analyzed, some basic insights can be formulated which should be kept in mind when considering the application of QA:

\begin{itemize}
    \item \textbf{Quantum annealing can only address specific problem formulations.} 
    Any problem intended to be solved by a practitioner with QA needs to be formulated in a specific manner targeted to the hardware architecture to be used. For D-Wave's superconducting QPUs, this amounts to formulating problems as Ising models or QUBO problems, where spin or binary variables are used to encode the problem. Currently, state-of-the-art quantum annealers can only encode linear and quadratic terms, and therefore all terms and constraints in the objective function must be implemented in that form. Furthermore, QPUs have fixed topologies of qubit layouts, which either restrict the connectivity of applicable problems, or require minor-embedding/hybrid techniques to overcome. 
    In short: while the QA algorithm is in theory intended to solve NP-hard problems, in practice, arbitrary problems may incur significant polynomial (or even exponential) overhead when transformed to the hardware-specific formulation.

    \item \textbf{Quantum annealing is a metaheuristic quantum optimization algorithm.}
    Although inspired by Adiabatic Quantum Computation, QA is implemented with more relaxed conditions, thus removing the theoretical guarantees of the adiabatic theorem in favor of near-term applicability. The result is a metaheuristic quantum optimization algorithm that exploits quantum effects for computation. There is reason to believe that quantum fluctuations are effective for transitions between spin configurations during the annealing, thus aiding global optimization; however, conclusive evidence has remained elusive despite rigorous research. For all practical purposes, QA is therefore a stochastic black-box optimization approach for combinatorial optimization, similar to simulated thermal annealing.
    
    \item \textbf{Only some problems are good candidates for quantum annealing.} Not all problems are currently good fits for QA, and it is difficult to know \emph{a priori} if any given real-world problem will benefit from QA. For example, performance on continuous optimization or high-precision problems suffers from transformations to QUBO/Ising due to noise and restricted qubit counts. Problems known to be in the computational complexity class $P$, which can be addressed efficiently with classical computing, may not be good candidates because QA is a metaheuristic with potentially unbounded worst-case performance. Similarly, decision problems suffer from the uncertainty of whether no feasible solution exists, or the solution simply hasn't been observed by the annealer. However, problems that did seem to exhibit good results with QA typically have simple but hard combinatorial properties and/or constraints where the number and connectivity of the qubits required for the problem do not scale strongly with the problem size. Therefore, while these guidelines offer some insights, more research is needed to understand the impact on real-world problems in general.

    \item \textbf{Hybrid algorithms are currently necessary.} Ising models or QUBO instances arising from real-world problems typically require more complicated logical variables, constraint terms, or interactions than can be realized with current hardware QA systems. This has inspired development of hybrid quantum-classical algorithms to overcome some of these limitations. Although there are many ways to construct such algorithms, for real-world problems reviewed in this article, problem partitioning is the most common approach. Problems are decomposed into sub-problems that are individually submitted to the QPU, with their respective results stitched together. Classical search and optimization routines are used to construct the outer loops in hybrid algorithms that orchestrate the sub-problem generation and solution recombination in an attempt to solve the original full problem. 
 
    \item \textbf{Quantum annealing can only reach its full potential through quantum hardware.} 
    Simulating quantum systems with classical hardware is in general inefficient and sometimes even intractable. There is evidence that QA implemented on D-Wave QPUs can simulate some quantum mechanical models more efficiently than classical algorithms. This, however, was strongly dependent on mapping intrinsic properties of the problem directly to some core functionality implemented in the quantum hardware (e.g., reverse annealing and pause/ramp of the transverse field).

\end{itemize}

\section{Conclusions}
\label{sec:conclusions}
Quantum annealing (QA) is a heuristic optimization algorithm used to solve problems in combinatorial optimization. It can be motivated by the idea of utilizing quantum fluctuations for an annealing process through which heuristic optimization is performed. Recently, advances in quantum technologies have enabled the development of quantum processors that implement the QA algorithm for programmable use in hardware.
The first commercially available hardware was provided by D-Wave Systems Inc.\ and  as a result, this platform is used most frequently to solve optimization problems across different disciplines. 
Quantum annealing processors -- in their current form -- have been developed specifically to be easier to build and operate compared to gate-based quantum computers. 
However, many questions and challenges still remain open, and continuous efforts are made to evaluate and extend QA hardware and its efficient utilization for addressing real-world application problems.
The ultimate question of whether QA offers a provable advantage over classical approaches has yet to be answered. 

In this paper, we first provided a general presentation of QA by explaining the foundations of this heuristic quantum optimization algorithm and the software and hardware characterizing present quantum processors. These concepts are essential to evaluate the potential and limitations of the method based on the state-of-the-art studies. 
We reviewed some example applications and empirical studies to provide the reader with updated information about the current characteristics and the applications status of this relatively new quantum technology, such as what problems can be addressed, what problem sizes are allowed, what is the expected scaling behavior, what constraints can be represented, etc.
The classification of this research was mainly organized according to the most relevant application fields.

In summary, based on all the works reviewed in this article, the empirical performance of the QA algorithm is highly dependent on the problem being addressed. The results are influenced by the interactions between the problem formulation, the encoding of the logical variables into qubits, and the embedding in the hardware platform. 
Regardless of the application field, only relatively small problem sizes can currently be addressed efficiently on actual hardware.
In addition to the small number of physical qubits, a major limiting factor is the overhead resulting from the restricted connection topology of the physical qubits  that is currently realizable in hardware. Typically, the number of physical qubits needed to address a problem scales polynomially with the number of logical variables in the problem formulation.  
However, after the careful analysis of a given problem class and its mapping to QA hardware, the hardware can be programmed for that problem class and then each instance of that problem can be solved robustly in a fixed amount of time, which is a major advantage of the quantum annealing approach. 

In order to take full advantage of QA's potential, it is clear that the unique properties of the QA hardware must be exploited in ways that classical software cannot compensate for. This indicates that novel features such as additional qubit control, driver Hamiltonians and operators, and higher connectivity must be developed and implemented in hardware for QA to be applicable in practice. The fast development of current and future hardware systems fuels the hope that exciting developments can be expected in the near to mid term and more realistic application problems can be addressed with quantum annealing.

\begin{acks}
The authors thank Markus Olhofer, Andrea Skolik, and Michele Cattelan for fruitful discussions and feedback. 
This work is part of the research programme C2D–Horizontal Data Science for Evolving Content with project name DACCOMPLI and project number 628.011.002, which is (partly) financed by the Netherlands Organisation for Scientific Research (NWO).
\end{acks}

\bibliographystyle{ACM-Reference-Format}


\end{document}